\documentclass[fleqn,usenatbib]{mnras}

\usepackage{newtxtext,newtxmath}

\usepackage[T1]{fontenc}

\DeclareRobustCommand{\VAN}[3]{#2}
\let\VANthebibliography\thebibliography
\def\thebibliography{\DeclareRobustCommand{\VAN}[3]{##3}\VANthebibliography}

\usepackage{graphicx}	
\usepackage{amsmath}	
\usepackage{enumitem}

\newcommand{\Omegab}{\Omega_\mathrm{b}}
\newcommand{\Omegai}{\Omega_\mathrm{i}}
\newcommand{\thetaf}{\theta_\mathrm{f}}
\newcommand{\thetas}{\theta_\mathrm{s}}
\newcommand{\thetasres}{\theta_\mathrm{s,res}}
\newcommand{\Jf}{J_\mathrm{f}}
\newcommand{\Js}{J_\mathrm{s}}

\newcommand{\ks}{k_\mathrm{s}}

\title[Bar resonances in the highly radial halo]{Radial halo substructure in harmony with the Galactic bar}

\author[A. M. Dillamore et al.]{
Adam M. Dillamore,$^{1}$\thanks{E-mail: amd206@cam.ac.uk (AMD)}
Vasily Belokurov,$^{1}$
and N. Wyn Evans$^{1}$
\\
$^{1}$Institute of Astronomy, University of Cambridge, Madingley Road, Cambridge CB3 0HA, UK\\
}

\date{Accepted XXX. Received YYY; in original form ZZZ}

\pubyear{2023}

\begin{document}
\label{firstpage}
\pagerange{\pageref{firstpage}--\pageref{lastpage}}
\maketitle

\begin{abstract}
Overdensities in the radial phase space $(r,v_r)$ of the Milky Way's halo have previously been associated with the phase-mixed debris of a highly radial merger event, such as \textit{Gaia} Sausage-Enceladus. We present and test an alternative theory in which the overdense `chevrons' are instead composed of stars trapped in resonances with the Galactic bar. We develop an analytic model of resonant orbits in the isochrone potential, and complement this with a test particle simulation of a stellar halo in a realistic barred Milky Way potential. These models are used to predict the appearance of action space $(J_\phi,J_r)$ and radial phase space in the Solar neighbourhood. They are able to reproduce almost all salient features of the observed chevrons. In particular, both the analytic model and simulation predict that the chevrons are more prominent at $v_r<0$ when viewed near the Sun, as is observed by \textit{Gaia}. This is inconsistent with formation by an ancient merger event. We also associate individual chevrons with specific resonances. At a bar pattern speed of $\Omegab=35$~km\,s$^{-1}$kpc$^{-1}$, the two most prominent prograde chevrons align very closely with the corotation and outer Lindblad resonances. The former can be viewed as a highly eccentric extension of the Hercules stream. Finally, our model predicts that the $v_r$ asymmetry changes sign as a function of Galactic radius and azimuth, and we find evidence that this is indeed the case in the Milky Way.
\end{abstract}

\begin{keywords}
Galaxy: kinematics and dynamics -- Galaxy: halo -- Galaxy: structure
\end{keywords}



\section{Introduction}

In the days of the old quantum theory, Max Born developed perturbation theory to cope with the effects of resonances. This is summarized in his 1926 book {\it`The Mechanics of the Atom'}. When there are commensurable frequencies in a dynamical problem, Born showed how to perform a canonical transformation to a new set of action-angle coordinates -- the so-called fast and slow action-angles. The basic idea is then to average over the rapid variations in the fast coordinates, keeping the slow variables fixed. Hamilton's equations then show that the fast action is constant or adiabatically invariant. This powerful technique was introduced into galactic dynamics by \citet{LB73} and exploited in the struggle to understand the origin of spiral structure and bar formation~\citep[e.g.,][]{LB79, Co79, Ea96, Co97}. It has also been widely used to model the capture of planets and planetesimals in solar system dynamics~\citep[e.g.,][]{Yo79, He82, Bo84}. 

The influence of the Galactic bar on stars in the stellar disc has long been realised to be substantial. \citet{Ka91} made the pioneering suggestion that, if the Sun were located near a resonance, then there may be two stellar streams, one moving inward and the other outward. Resonant capture is an important mechanism for generating substructure in the stellar disc~\citep[e.g.,][]{De00,sellwood2010,Ge11,Mc13,binney2020,chiba2021,chiba2021_treering}. With the advent of {\it Gaia} data, the ridges and features in the phase space distribution of disc stars have been subjected to scrutiny and often attributed to bar or spiral arm resonances~\cite[e.g.,][]{Fr19,khoperskov2020,trick2021,Tr22,Kh22,wheeler2022}. 

By contrast, the effect of bar resonances in the halo has not received as much attention. \citet{Deb98, De00} demonstrated that dynamical friction from dark matter haloes acts to spin down bars. The phenomenon of dynamical friction is driven by the trapping of dark matter particles by the bar. This showed that resonances due to bars affect the dynamics of stars and dark matter in the halo, though the emphasis of the work was on implications for bar evolution and dark matter density. This has been further explored by \citet{athanassoula2002,We02,Ce07,collier2019,chiba2022}; and \citet{hamilton2023}. The possibility of bar resonances creating substructures in the stellar halo was first suggested by \citet{moreno2015, moreno2021}. They showed that the trapping can extend several kpc above or below the Galactic plane and suggested that it may be the origin of some known moving groups. The first demonstration that this effect is present in the data on stellar halo stars was provided in \citet{My18} who noted the presence of a possible resonance in halo stars at all metallicities.

The \textit{Gaia} Data Release 3 \citep[DR3;][]{gaia_dr3} provided a large sample of halo stars with full phase space coordinates around the Sun. \citet{belokurov_chevrons} exploited this to identify prominent chevrons in the space of radial position and velocity $(r,v_r)$. These are the tell-tale signatures of shells, which are known to form in nearly radial mergers~\citep[e.g.,][]{filmore1984,Quinn84,Am17,dong-paez2022,davies2023_ironing}. A natural conclusion was that the chevrons were caused by shells, expected from the last significant merger of the Milky Way with \textit{Gaia} Sausage-Enceladus~\citep[GSE;][]{belokurov2018,helmi2018}. 

Though natural, this conclusion does not appear to be correct. First doubts emerged with the realisation that the chevrons are present in comparatively metal-rich stars, whereas the dwarf progenitor is expected to be metal-poor. Only a very small fraction of the GSE stars have [Fe/H] $> -0.7$~\citep{naidu2020,feuillet2021}. The chevrons are also highly asymmetric in $v_r$, being clearly visible only at $v_r<0$. \citet{zhang2023_vmp} similarly noted that metal-poor stars on highly radial orbits are distributed asymmetrically in velocity space, with a preference for $v_r<0$. \citet{donlon2023} showed that the asymmetry of the chevrons is inconsistent with them being dynamically old, and it is therefore difficult to explain how they could have been formed by the ancient GSE. They argued that this provides evidence of a more recent radial merger event in the last few Gyr, dubbed the Virgo Radial Merger (VRM).

However, the chevrons may not have been created by any merger event. \citet{dillamore2023} demonstrated that similar structures are generated in simulations by the trapping of particles in resonances with the bar, with the corotation resonance being particularly prominent. Using a bar pattern speed in the range $\Omega_{\rm b} \approx 35-40$ km s$^{-1}$ kpc$^{-1}$ provides a good match both to the substructures in phase space and the measured speed of the Galactic bar~\citep{Sa19,binney2020,chiba2021_treering}. \citet{davies2023_bar} also showed that non-resonant merger-generated chevrons may be destroyed by the rotating bar, raising doubts about their survivability in the Milky Way.

The purpose of this paper is to test the hypothesis that the chevron stars are trapped in bar resonances. In Section~\ref{section:model}, we develop an analytic model of resonances in the isochrone potential. We then describe and present the results of our test particle simulation in Section~\ref{section:simulations} and \textit{Gaia} data in Section~\ref{section:data}. We analyse and compare the analytic model, simulation and data in Section~\ref{section:analysis} and present our conclusions in Section~\ref{section:conclusions}. Finally, Appendices~\ref{section:orbits_appendix}, \ref{section:G_appendix} and \ref{section:sim_appendix} describe aspects of the analytic model and simulation in more detail.

\section{Analytic model}\label{section:model}

In this section, we provide an analytic model of resonances with the bar. While this cannot compete with the realism of simulations, it will allow us to build understanding of the dynamics involved, and make qualitative predictions to be compared with simulations and observations.

We treat the overall potential as the sum of a static spherical background potential $\Phi_0$ and a small time-dependent non-axisymmetric perturbation $\Phi_\mathrm{b}$ due to the bar:
\begin{equation}\label{eq:potential}
    \Phi(r,\theta,\phi,t)=\Phi_0(r)+\Phi_\mathrm{b}(r,\theta,\phi,t),
\end{equation}
where $(r,\theta,\phi)$ are Galactocentric spherical coordinates. In the unperturbed potential $\Phi_0$, we may describe the orbits in terms of angle-action variables $(\boldsymbol{\theta}, \boldsymbol{J})$. The actions $\boldsymbol{J}$ are integrals of motion, so the unperturbed Hamiltonian can be expressed as a function of the actions only, $H_0(\boldsymbol{J})$. Hamilton's equations then read
\begin{align}
    \dot{J_i}&=-\frac{\partial H}{\partial \theta_i}=0\\
    \dot{\theta}&=\frac{\partial H}{\partial J_i}\equiv\Omega_i(\boldsymbol{J}).\label{eq:frequencies}
\end{align}
The angles $\boldsymbol{\theta}$ therefore evolve linearly in time. They are scaled such that one period corresponds to $\theta_i$ increasing by $2\pi$, so that $\Omega_i$ are the fundamental frequencies of motion \citep{binney_tremaine}. In the perturbed potential $\Phi$, the actions $\boldsymbol{J}$ are no longer exact integrals of motion and the true frequencies are not equal to $\Omega_i$. However, in the limit $|\Phi_\mathrm{b}/\Phi_0|\to0$ the true frequencies tend to $\Omega_i$ \citep{trick2021}. We use this weak bar limit in our model, so that $\Omega_i$ are good approximations to the true frequencies.

We consider resonances of the form
\begin{equation}\label{eq:resonances}
    m(\Omega_\phi-\Omegab)+l\,\Omega_r=0,
\end{equation}
where $\Omega_\phi$ and $\Omega_r$ are the azimuthal frequency and the radial frequency of oscillation respectively, and $l$ and $m$ are integers. We take the bar's pattern speed to be positive, $\Omegab>0$. We emphasise that we use this equation to define both prograde and retrograde resonant orbits (i.e. both positive and negative $\Omega_\phi$)\footnote{This differs from the convention used by \citet{Co97}, in which the resonance condition was changed for retrograde orbits so that the formulae for the $l/m$ resonant orbit at $J_\phi > 0$ match onto the $l/m + 1$ resonant orbit at $J_\phi < 0$ (see Section~\ref{section:retrograde}).}. Studies of resonances in the disc frequently employ the epicyclic approximation \citep{binney_tremaine} to calculate $\Omega_\phi$ and $\Omega_r$ for a perturbed circular orbit in a given potential. However, we wish to consider highly eccentric orbits, far from where this approximation is valid. In order to develop an analytic model of the resonances, we therefore wish to work in a potential in which equation~\ref{eq:frequencies} can be evaluated for all $\boldsymbol{J}$. \citet{evans1990} showed that the most general potential for which the actions can be written in terms of elementary functions is the isochrone potential,
\begin{equation}\label{isochrone}
    \Phi_\mathrm{I}(r)=-\frac{GM}{b+\sqrt{b^2+r^2}},
\end{equation}
where $M$ is the total mass and $b$ is a scale length parameter. In the limit $r\to0$ the potential tends to the harmonic limit of a homogeneous sphere, and when $r\to\infty$ it tends to the Keplerian limit of a point mass $M$. While $\Phi_\mathrm{I}$ is too simple for modelling data in the Milky Way (for example, it is spherically symmetric), it has the enormous benefit that all orbits are fully analytic \citep{Henon, ELS,binney_tremaine}. It is therefore an excellent tool for developing physical understanding and making qualitative predictions. We therefore choose our unperturbed potential to be $\Phi_0(r)=\Phi_\mathrm{I}(r)$.

\begin{figure}
  \centering
  \includegraphics[width=\columnwidth]{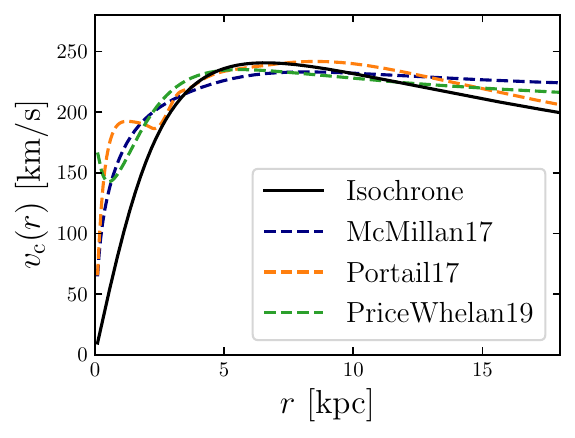}
  \caption{Circular velocity $v_\mathrm{c}(r)$ of the isochrone potential for our choice of parameters, with three realistic Milky Way models for comparison. The mass $M$ of the potential is chosen such that $v_\mathrm{c}$ matches the observed value in the Solar neighbourhood, and the selected scale radius $b$ gives a reasonably flat rotation curve across most of the Galactic disc. Our isochrone model is a good approximation to the Milky Way for $r\gtrsim4$~kpc, but the less concentrated central mass distribution results in a smaller $v_\mathrm{c}(r)$ at $r\lesssim4$~kpc.} 
   \label{fig:rot_curve}
\end{figure}

The circular velocity in the isochrone potential obeys
\begin{equation}
    v_\mathrm{c}^2(r)=\frac{GMr^2}{(b+a)^2a},\qquad\qquad
    a\equiv\sqrt{b^2+r^2}.
\end{equation}
In order to match the circular velocity of the Milky Way $v_0$ at the Sun's radius $r_0$, we choose
\begin{align}
    GM=\frac{v_0^2}{r_0^2}\left(b+\sqrt{b^2+r_0^2}\right)^2\sqrt{b^2+r_0^2}.
\end{align}
Following \citet{Po17}, we set $r_0=8.2$~kpc and $v_0=238$~km\,s$^{-1}$, and choose $b=3$~kpc. This gives $M=2.35\times10^{11}M_\odot$ and a maximum circular velocity of $v_\mathrm{max}\approx241$~km\,s$^{-1}$. The rotation curve for this choice of parameters is shown in Fig.~\ref{fig:rot_curve}. For comparison, we also show several realistic potentials fitted to data from the Milky Way, namely those of \citet{mcmillan17} and \citet{Po17}, and the \textsc{MilkyWayPotential} from \textsc{Gala} \citep{gala}. The isochrone rotation curve is a reasonable match to the Milky Way at $r\gtrsim4$~kpc, but is less realistic at smaller radii. This is partly because the isochrone potential has a homogeneous core, while cosmological dark halos have a density cusp.
The circular velocity in the central regions is therefore smaller in $\Phi_\mathrm{I}$ than in the Milky Way. We set the pattern speed of the bar to $\Omegab=35$~km\,s$^{-1}$kpc$^{-1}$, close to many recent estimates for the Milky Way \citep[e.g.][]{Sa19,binney2020,chiba2021_treering,zhang24}. With this choice of parameters the corotation radius (where $v_\mathrm{c}(r)=r\Omegab$) is $r_\mathrm{CR}=6.87$~kpc. Throughout this paper we place the Sun at an angle relative to the bar of $\phi_\odot-\phi_\mathrm{b}=-30^\circ$, consistent with the measurements of e.g. \citet{wegg2015}.

\begin{figure}
  \centering
  \includegraphics[width=\columnwidth]{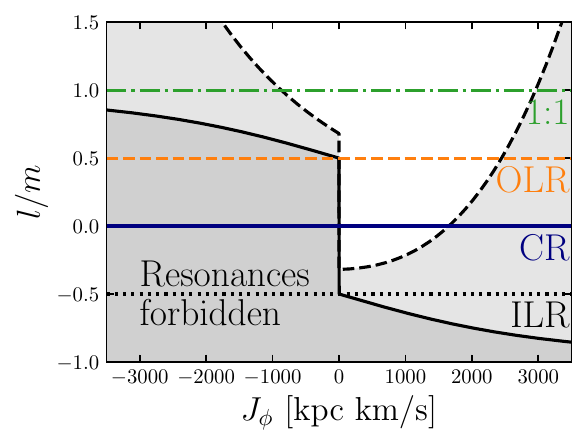}
  \caption{Minimum value of $l/m$ required for the existence of resonant orbits in the $\theta=\pi/2$ plane as a function of $J_\phi$, in the isochrone potential. Resonances are possible above the black dashed line for our choice of pattern speed $\Omegab$, and possible for some smaller value of $\Omegab$ above the black solid line. Below the black solid line they are impossible for this potential in this plane. The horizontal lines mark the positions of some of the most important resonances, including the corotation (CR) and outer Lindblad (OLR) resonances. The CR and OLR are always only possible on prograde orbits ($J_\phi>0$), but higher values of $l/m$ are possible for $J_\phi<0$. The inner Lindblad resonance (ILR) is not possible for this choice of potential and pattern speed. See Section~\ref{section:retrograde} for a discussion of retrograde resonant orbits.} 
   \label{fig:resonance_condition}
\end{figure}

\subsection{Resonances in the isochrone potential}

In this section we derive the loci of bar resonances in spaces of energy $E$ and radial action $J_r$ versus azimuthal action $J_\phi$. We utilise the property of the isochrone potential that the Hamiltonian and all actions and frequencies can be written as analytic functions of each other. As discussed above, this requires the approximation that the bar potential is a small perturbation, such that the true frequencies can be approximated by the unperturbed frequencies $\Omega_i$.

Following \citet{binney_tremaine}, we express the actions in spherical coordinates $(r,\theta,\phi)$. The unperturbed Hamiltonian is
\begin{align}
    H_0(\boldsymbol{J})&=-\frac{(GM)^2}{2\left[J_r+\frac{1}{2}\left(L+\sqrt{L^2+4GMb}\right)\right]^2},\\
    L&\equiv J_\theta+|J_\phi|.
\end{align}
$J_\phi=L_z$ and $L$ are the $z$-component and magnitude of angular momentum respectively. Using equation~\ref{eq:frequencies} the frequencies are then
\begin{align}
    \Omega_r&=\frac{(GM)^2}{\left[J_r+\frac{1}{2}\left(L+\sqrt{L^2+4GMb}\right)\right]^3},\label{eq:Omega_r}\\
    \Omega_\theta&=\frac{1}{2}\left(1+\frac{L}{\sqrt{L^2+4GMb}}\right)\Omega_r,\label{eq:Omega_theta}\\
    \Omega_\phi&=\mathrm{sgn}(J_\phi)\Omega_\theta.\label{eq:Omega_phi}
\end{align}
Substituting the above frequencies into the resonance condition (equation~\ref{eq:resonances}) gives~\citep[e.g.,][]{LB79, Ea96}
\begin{equation}
    \left[\frac{1}{2}\left(1+\frac{L}{\sqrt{L^2+4GMb}}\right)\mathrm{sgn}(J_\phi)+\frac{l}{m}\right]\Omega_r=\Omegab.
\end{equation}
Noticing that the Hamiltonian can be written as $H_0(\boldsymbol{J})=-\frac{1}{2}(GM\Omega_r)^{2/3}$, we can eliminate $\Omega_r$ to obtain the energy of the $(l,m)$ resonance as a function of $J_\phi$ and $L$. We can also find the radial action $J_r$. From equation~\ref{eq:Omega_r}, this is
\begin{align}
    J_r=\left[\frac{(GM)^2}{\Omega_r}\right]^\frac{1}{3}-\frac{1}{2}\left(L+\sqrt{L^2+4GMb}\right).
\end{align}
The energy and radial action along the $(l,m)$ resonance are therefore:
\begin{align}
    E_{(l,m)}&=-\frac{1}{2}(GM\Omegab)^\frac{2}{3}\left[\frac{1}{2}\left(1+\frac{L}{\sqrt{L^2+4GMb}}\right)\mathrm{sgn}(J_\phi)+\frac{l}{m}\right]^{-\frac{2}{3}},\label{eq:E_Lz_res}\\
    J_{r,(l,m)}&=\left[\frac{(GM)^2}{\Omegab}\right]^\frac{1}{3}\left[\frac{1}{2}\left(1+\frac{L}{\sqrt{L^2+4GMb}}\right)\mathrm{sgn}(J_\phi)+\frac{l}{m}\right]^\frac{1}{3}\label{eq:Jr_Jphi_res}\\&\quad-\frac{1}{2}\left(L+\sqrt{L^2+4GMb}\right),\notag
\end{align}
subject to the conditions,
\begin{align}
    \frac{l}{m}&>-\frac{1}{2}\left(1+\frac{L}{\sqrt{L^2+4GMb}} \right)\mathrm{sgn}(J_\phi),\label{eq:Omega_r_condition}\\
    \frac{l}{m}&\geq\frac{\Omegab}{(GM)^2}\left[\frac{1}{2}\left(L+\sqrt{L^2+4GMb}\right)\right]^3\label{eq:J_r_condition}\\
    &\qquad-\frac{1}{2}\left(1+\frac{L}{\sqrt{L^2+4GMb}}\right)\mathrm{sgn}(J_\phi).\notag
\end{align}
The first condition is required so that $\Omega_r>0$, and applies equally for all bar pattern speeds. The second ensures that $J_{r,(l,m)}\geq0$, and depends on the pattern speed $\Omegab$. If equations~(\ref{eq:Omega_r_condition}) and (\ref{eq:J_r_condition}) are not satisfied for particular choices of $J_\phi$, $L$ and $\Omegab$, then the $(l,m)$ resonant orbit cannot exist in our isochrone potential for those values. Note that for $\Omegab>0$, equation~\ref{eq:Omega_r_condition} is always satisfied if equation~(\ref{eq:J_r_condition}) is satisfied. The first condition represents the hard minimum of $l/m$ in the limit $\Omegab\to0$.

In Fig.~\ref{fig:resonance_condition}, we plot the minimum $l/m$ for resonance against $J_\phi$ for orbits in the $\theta=\pi/2$ plane. The solid and dashed black lines indicate the constraints from equations (\ref{eq:Omega_r_condition}) and (\ref{eq:J_r_condition}) respectively. For the outer Lindblad resonance (OLR, $l/m=1/2$) and below, resonant orbits are only ever possible for $J_\phi>0$. However, resonances are still possible for some retrograde orbits at higher $l/m$.

\begin{figure}
  \centering
  \includegraphics[width=\columnwidth]{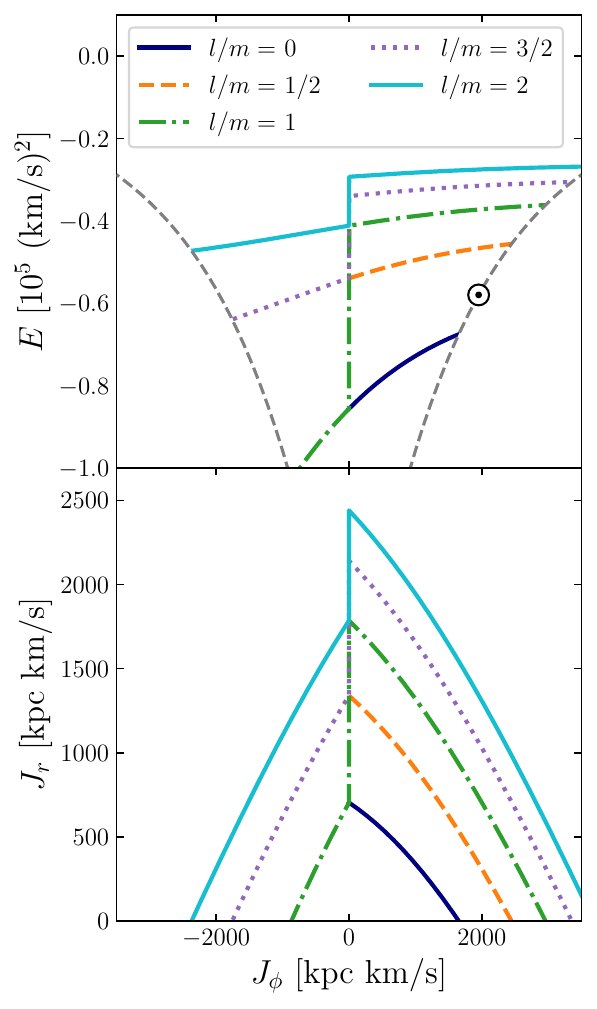}
  \caption{\textbf{Top panel:} Energy $E$ vs z-angular momentum $L_z=J_\phi$ for a series of resonances. Each colour corresponds to a different resonance as marked in the legend. The grey dashed lines mark the loci of circular orbits in the $\theta=\pi/2$ plane, with a circular orbit at the Sun's radius $r_0$ marked with the $\odot$ symbol. \textbf{Bottom panel:} As above, but showing radial action $J_r$ vs $J_\phi$.} 
   \label{fig:E_Lz}
\end{figure}

Using equations~(\ref{eq:E_Lz_res}) and (\ref{eq:Jr_Jphi_res}), we show the locations of a series of resonances in energy vs angular momentum space $(J_\phi,E)$ and action space $(J_\phi, J_r)$ in Fig.~\ref{fig:E_Lz}. These are the most important resonances, with $m=2$ and $l\in\{0,1,2,3,4\}$. This set includes the corotation resonances (CR, $l/m=0$) and the outer Lindblad resonance (OLR, $l/m=1/2$). For our choice of parameters, the inner Lindblad resonance (ILR, $l/m=-1/2$) does not exist in the isochrone potential. However, it does exist for smaller values of $\Omegab$, when the black dashed line in Fig.~\ref{fig:resonance_condition} would be at lower $l/m$. As $J_\phi$ decreases from its maximum value, the energy of each resonance decreases. This is because at fixed energy $\Omega_\phi$ decreases as $J_\phi$ decreases. Equation \ref{eq:resonances} dictates that in order to remain in the resonance, $\Omega_r$ must therefore increase to compensate. Hence, $E$ decreases as $J_\phi$ is reduced along each resonance. In $(J_r, J_\phi)$ space, the resonances are approximately straight lines for small $J_r$, as previously shown \citep[e.g.,][]{Co97,sellwood2010,trick2021}. 

\begin{figure}
  \centering
  \includegraphics[width=\columnwidth]{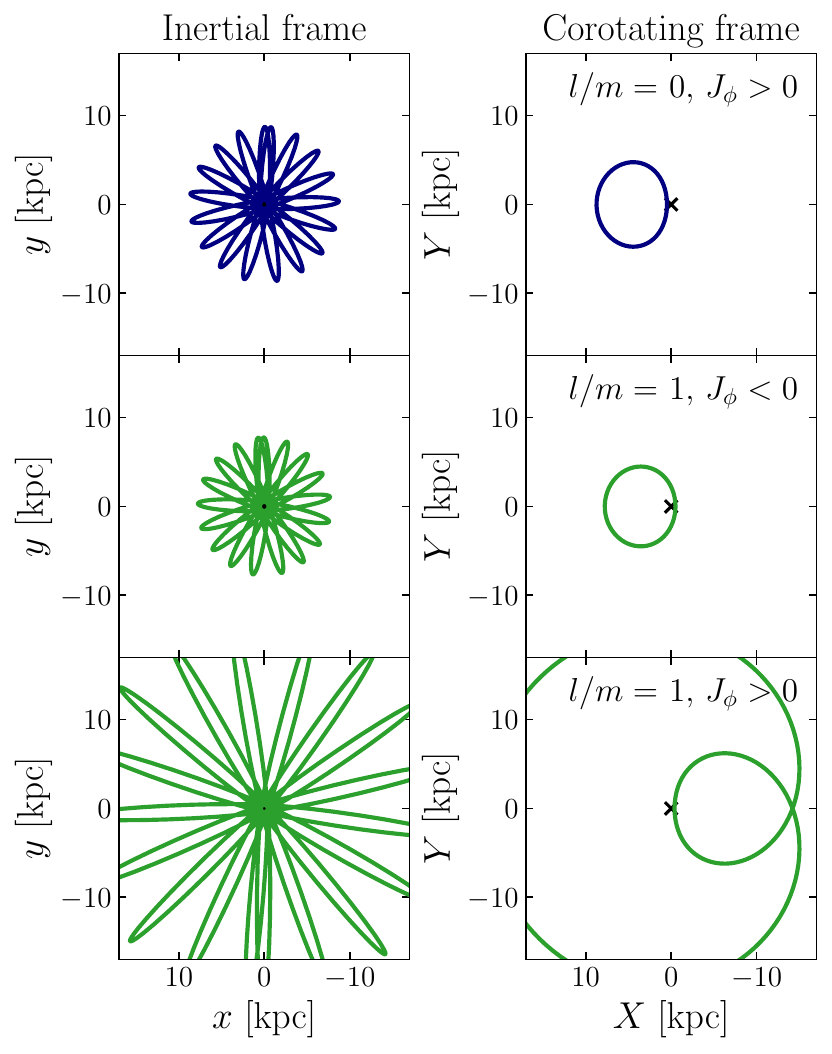}
  \caption{Shapes of the $l/m=0$ and $l/m=1$ resonant orbits with $|J_\phi|=200$~kpc km\,s$^{-1}$, in the inertial frame (left-hand column) and the frame corotating with the bar (right-hand column). The $l/m=1$ resonance is shown for both $J_\phi<0$ (middle row) and $J_\phi>0$ (bottom row). Note that the $X$-axis is flipped, such that the bar rotates in a clockwise sense. The Galactic centre is marked by the black crosses. The orientations of the orbits in the $(X, Y)$ frame are arbitrarily chosen.} 
  \label{fig:cr_orbits}
\end{figure}

\subsection{Retrograde orbits}
\label{section:retrograde}

The equations above are all valid for both prograde and retrograde orbits. However, the orbit on a given $(l,m)$ resonance changes discontinuously as the sign of $J_\phi$ is changed. Each prograde resonance is continuous in $E$ and $J_r$ with a retrograde resonance with a different value of $l$. For example, as expected from equation~\ref{eq:Omega_r_condition} and Fig.~\ref{fig:resonance_condition}, the CR and OLR exist only at $J_\phi>0$, but are continuous with higher resonances at $J_\phi<0$. Generally, each $l/m$ resonant orbit at $J_\phi>0$ is related to the $l/m+1$ resonant orbits at $J_\phi<0$. The continuity between these resonances on each side of $J_\phi=0$ was noted earlier \citep[e.g.,][]{kalnajs1997,Co97}, and can be understood intuitively by plotting the orbits in configuration space. In Appendix~\ref{section:orbits_appendix}, we give equations relating the angle variables $\theta_i$ to the physical coordinates of particles on given orbits $(r,\phi)$. We use these equations to plot the orbits of the $l/m=0$ and $l/m=1$ resonances for $J_\phi=\pm200$~kpc km\,s$^{-1}$ in Fig.~\ref{fig:cr_orbits}. The left-hand column shows the inertial frame, and the right-hand column shows the frame co-rotating with the bar (i.e. rotating at frequency $\Omegab$). 

The orbits in the inertial frame are all qualitatively similar, consisting of highly eccentric motion around a curve whose apocentres and pericentres precess by some angle per radial period. The apocentric radii of the prograde $l/m=0$ and retrograde $l/m=1$ orbits are similar, while that of the prograde $l/m=1$ orbit is much greater, as expected from the higher energy.

The right-hand column is more informative. Let $\phi'$ be the angle measured from the $X$-axis in this frame (i.e. $\mathrm{tan}\,\phi'=Y/X$). For the CR ($l/m=0$) orbit, notice the Galactic centre is not enclosed by the orbit, so $\phi'$ oscillates about a certain value instead of circulating through all possible values. This is why the orbit is in corotation, with $\Omega_\phi=\Omegab$. Now consider reducing the angular momentum $J_\phi$ of the orbit. The pericentre decreases until it reaches zero at $J_\phi=0$. If $J_\phi$ continues to become negative, the pericentre crosses to the opposite side of the origin, such that the Galactic centre is now enclosed. This is the case seen in the middle row. Now $\phi'$ circulates between $0$ and $2\pi$ in the same time as one radial period. Therefore $\Omega_\phi-\Omegab=\Omega_r$, and $l/m=1$. In the same way any prograde $l/m$ resonant orbit matches onto a retrograde $l/m+1$ orbit.

Meanwhile, the prograde $l/m=1$ orbit also circulates in $\phi'$ once per radial period, but the frequencies are considerably slower. This is because the orbital motion is in the same sense as the bar's rotation, so the magnitude of the relative frequency $|\Omega_\phi-\Omegab|$ is smaller.

\subsection{Particle dynamics near resonance}\label{section:pendulum}

So far we have neglected the effects of the bar itself, considering only the exact resonances of an infinitesimal perturbation rotating with some pattern speed. We must now address its influence on orbits close to the resonances, as this will determine their stability. This can tell us, for example, which orientations of resonant orbits relative to the bar are possible. We follow the method outlined by \citet{Tr84} and others \citep[e.g.][]{binney2018,chiba2021,hamilton2023}, considering a steadily rotating bar with pattern speed $\Omegab$. Following \citet{chiba2021}, we set the potential of the bar perturbation in the $\theta=\pi/2$ plane to
\begin{align}
    \Phi_\mathrm{b}(r, \phi, t)&=\Phi_2(r)\,\mathrm{cos}\left[2(\phi-\Omegab t)\right],\\
    \Phi_2(r)&=-A\frac{r^2}{\left[r_\mathrm{b}+r\right]^5},\label{eq:Phi_2}
\end{align}
where $r_\mathrm{b}$ is the scale length of the bar and $A$ is a positive constant. This is a quadrupole perturbation with an azimuthal minimum lying parallel to $\phi=\Omegab t\equiv\phi_\mathrm{b}$, the major axis of the bar. We set $r_\mathrm{b}=1.9$~kpc. This is approximately equal to $0.28\,r_\mathrm{CR}$, following \citet{chiba2021} and \citet{hamilton2023}.

For a particle orbiting close to the $(l,m)$ resonance, define the slow angle,
\begin{align}
    \thetas&\equiv l\theta_r+m(\theta_\phi-\phi_\mathrm{b})\label{eq:theta_s}\\
    &=l\theta_r+m(\theta_\phi-\Omegab t),
\end{align}
which evolves with frequency
\begin{equation}
\Omega_\mathrm{s}\equiv\dot{\theta}_\mathrm{s}=l\Omega_r+m(\Omega_\phi-\Omegab).
\end{equation}
We see from equation~\ref{eq:resonances} that near the $(l,m)$ resonance this frequency will be much smaller than $\Omega_\phi$ and $\Omega_r$. We also define the fast angle $\theta_\mathrm{f}$, choosing it to be equal to the radial angle:
\begin{equation}
    \thetaf\equiv\theta_r.
\end{equation}
We wish to make a canonical transformation from the original angle-action variables $(\boldsymbol{\theta},\boldsymbol{J})$ to a new set $(\boldsymbol{\theta}',\boldsymbol{J}')$, where $\boldsymbol{\theta}'\equiv(\thetaf,\thetas)$ and $\boldsymbol{J}'\equiv(\Jf,\Js)$. This will allow us to separate the orbital motion into fast ($\thetaf,\Jf$) and slow ($\thetas,\Js$) components.

We use the generating function~\citep[e.g.,][]{LB73,Co97}
\begin{align}
    W(\boldsymbol{\theta},\boldsymbol{J},t)=\left[l\theta_r+m(\theta_\phi-\Omegab t)\right]\Js+\theta_r\Jf,
\end{align}
which satisfies one equation required for a canonical transformation
\begin{equation}
    \boldsymbol{\theta}'=\frac{\partial W}{\partial \boldsymbol{J}'}.
\end{equation}
The other equations describing the transformation then tell us
\begin{align}
    \boldsymbol{J}\equiv(J_r,J_\phi)&=\frac{\partial W}{\partial \boldsymbol{\theta}},\\
    &=(l\Js+\Jf,m\Js),\\
H'(\boldsymbol{\theta}',\boldsymbol{J}',t)&=H(\boldsymbol{\theta},\boldsymbol{J},t)+\frac{\partial W}{\partial t}\\
&=H(\boldsymbol{\theta},\boldsymbol{J},t)-m\Omegab\Js,
\end{align}
where $H'$ is the perturbed Hamiltonian. The slow and fast actions are therefore $\Js=J_\phi/m$ and $\Jf=J_r-\frac{l}{m}J_\phi$. We split the Hamiltonian into the unperturbed stationary part and a perturbation due to the bar, which we expand as a Fourier series:
\begin{align}
H'(\boldsymbol{\theta}',\boldsymbol{J}')&=H_0(\boldsymbol{J}')+\sum_{\boldsymbol{k}}\Psi_{\boldsymbol{k}}e^{i\boldsymbol{k}\cdot\boldsymbol{\theta}'}-m\Omegab\Js,\\
\boldsymbol{k}&\equiv(k_\mathrm{f},k_\mathrm{s}).
\end{align}
The Fourier coefficients $\Psi_{\boldsymbol{k}}$ can be expressed in terms of the potential of the bar perturbation $\Phi_\mathrm{b}$ \citep[see Appendix B of][]{chiba2021}. The time dependence of the Hamiltonian has been absorbed into the slow angle $\thetas$ \citep{hamilton2023}. Since $\thetaf$ evolves much faster than $\thetas$, we may average over $\thetaf$ to obtain:
\begin{align}
\bar{H'}(\thetas,\boldsymbol{J}')&=H_0(\boldsymbol{J}')+\sum_{\ks}\Psi_{\ks}e^{i\ks\thetas}-m\Omegab\Js,\\
\Psi_{\ks}&\equiv\Psi_{(0,\ks)}.
\end{align}
For the important resonances for which $m=2$ (including all those shown in Fig.~\ref{fig:E_Lz}), only $\ks=\pm1$ terms contribute to the sum \citep{chiba2021}. The Hamiltonian must also be real, implying that
\begin{equation}
    \bar{H'}(\thetas,\boldsymbol{J}')=H_0(\boldsymbol{J}')+2\Psi_1(\boldsymbol{J}')\,\mathrm{cos}\,\thetas-m\Omegab\Js,
\end{equation}
where we have used the fact that $\Psi_1$ is real, as shown by \citet{chiba2021}:
\begin{equation}
    \Psi_1(\boldsymbol{J}')=\frac{1}{2\pi}\int_0^\pi\mathrm{d}\theta_r\Phi_2(r)\,\mathrm{cos}\left[2\left(\phi-\theta_\phi-\frac{l}{m}\theta_r\right)\right].\label{eq:Psi_1}
\end{equation}
Hamilton's equations then give~\citep[e.g.,][]{LB79, Co97}
\begin{align}
    \dot{J}_\mathrm{s}&=-\frac{\partial\bar{H'}}{\partial\thetas}=2\Psi_1\mathrm{sin}\,\thetas,\label{eq:J_s_dot}\\
    \dot{\theta}_\mathrm{s}&=\frac{\partial\bar{H'}}{\partial\Js}=\frac{\partial H_0}{\partial \Js}+2\frac{\partial \Psi_1}{\partial \Js}\mathrm{cos}\,\thetas-m\Omegab.\label{eq:theta_s_dot}
\end{align}
We differentiate equation~\ref{eq:theta_s_dot} and substitute in equation~\ref{eq:J_s_dot}. Discarding small terms containing $O(\Psi_1^2)$ or $\dot{\theta}_\mathrm{s}$ gives the pendulum equation \citep[e.g.,][]{LB73,chirikov1979,Co97},
\begin{align}   \ddot{\theta}_\mathrm{s}&=2G\Psi_1\mathrm{sin}\,\thetas,\label{eq:pendulum}\\
    G&\equiv\frac{\partial^2H_0}{\partial\Js^2}.\label{eq:G}
\end{align}
The quantity $G$ is sometimes called the `cooperation parameter' \citep{Ea96} or the `inertial response' \citep{Co97}.

Equation~\ref{eq:pendulum} shows that $\thetas$ can either \textit{circulate} through a full range of $2\pi$ or \textit{librate} (oscillate) about a stable equilibrium point $\thetasres$. We are interested in the implications of this equation for the stability of librating resonant orbits. The value of $\thetasres$ depends on the sign of $G\Psi_1$:
\begin{equation}
    \thetasres=
    \begin{cases}
        0, & G\Psi_1<0\\
        \pi, & G\Psi_1>0.
    \end{cases}
\end{equation}
We can relate $\thetasres$ to the orientations of the stable resonant orbits by evaluating equation~\ref{eq:theta_s} at a pericentre, where $\theta_r=0$ and $\theta_\phi=\phi$. Hence $\thetasres=m(\phi-\phi_\mathrm{b})$. For the $m=2$ resonances this gives us the locations of the stable pericentres relative to the bar:
\begin{equation}
    \phi_\mathrm{peri}-\phi_\mathrm{b}=
    \begin{cases}
        0, \pi, & G\Psi_1<0\\
        \pm\frac{\pi}{2}, & G\Psi_1>0.
    \end{cases}
\end{equation}
We calculate $G$ analytically for the isochrone potential in Appendix~\ref{section:G_appendix}. For most of the orbits we consider, $G<0$. However, the sign of $\Psi_1$ depends on the orbit. For example, consider a prograde low-eccentricity orbit at the corotation resonance, $l=0$. Near a circular orbit $\phi-\theta_\phi$ is small, so the cosine in the integrand of equation~\ref{eq:Psi_1} will always be positive. Since $\Phi_2(r)<0$ by definition (equation~\ref{eq:Phi_2}), $\Psi_1<0$ for this orbit. Therefore $\phi_\mathrm{peri}-\phi_\mathrm{b}=\pm \pi/2$, and the orbit will be centred on the bar's $L_4$ and $L_5$ Lagrange points. This is the result expected for low-eccentricity orbits \citep[e.g. see Section 3.3.2 in][]{binney_tremaine}.

Now consider a very highly eccentric prograde orbit, still with $l=0$. As the particle moves away from its pericentre, $\phi$ will increase by $\sim\pi/2$ very rapidly as the orbit passes close to the origin. However, $\theta_\phi$ will increase much more slowly, at a uniform rate throughout the orbit. This will result in $\phi-\theta_\phi$ being closer to $\pi/2$ than $0$, so the cosine will be negative for this part of the orbit. Since the weighting by $\Phi_2(r)$ is greatest for small $r$, this section of the orbit may be able to dominate the integral and cause $\Psi_1$ to be positive. In this case $\phi_\mathrm{peri}-\phi_\mathrm{b}=0$ or $\pi$. For this to occur, the bar's scale length $r_\mathrm{b}$ must be sufficiently short that the small-$\theta_r$ part of the integral dominates. Similar arguments can be applied to higher resonances with $l>0$.

We therefore expect that, depending on the length of the bar, the angles of the resonant orbits' pericentres may flip from the bar's minor axis to its major axis as their angular momentum is reduced. We can test this prediction by using our analytic expression for $G$ (equation~\ref{eq:G_analytic}) and numerically integrating equation~\ref{eq:Psi_1} for a range of orbits. We use the equations in Appendix~\ref{section:orbits_appendix} to convert between the angles $(\theta_r,\theta_\phi)$ and the physical position $(r,\phi)$.

\begin{figure}
  \centering
  \includegraphics[width=\columnwidth]{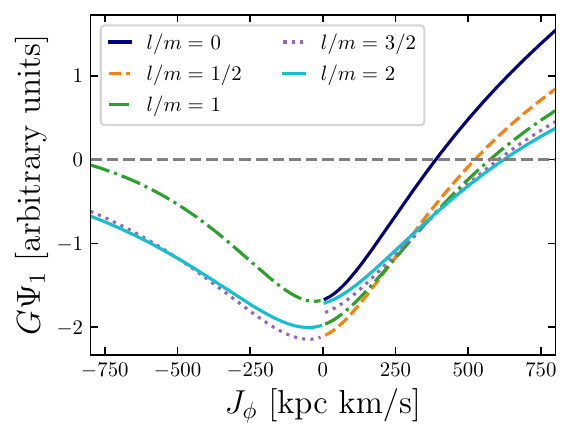}
  \caption{$G\Psi_1$ as a function of $J_\phi$ along the same series of resonances as Fig.~\ref{fig:E_Lz}. As $J_\phi$ decreases below $\approx400-600$~kpc km\,s$^{-1}$, $G\Psi_1$ becomes negative. This results in the locations of the stable pericentres switching to the major axis of the bar.} 
  \label{fig:GPsi_1}
\end{figure}

For the resonances shown in Fig.~\ref{fig:E_Lz}, we calculate the product $G\Psi_1$ across a grid of orbits in the $\theta=\pi/2$ plane with a range of $J_\phi$. The results are shown in Fig.~\ref{fig:GPsi_1}.

The behaviour of $G\Psi_1$ is generally in line with the above predictions. For large positive $J_\phi$, $G\Psi_1>0$ and therefore $\phi_\mathrm{peri}-\phi_\mathrm{b}=\pm\pi/2$. This corresponds to the orbits being centred on the $L_4$ or $L_5$ Lagrange points. However, when $J_\phi\lesssim500$~kpc km\,s$^{-1}$ the sign of $G\Psi_1$ changes to negative. Hence for highly eccentric and retrograde resonant orbits, the stable pericentres are at $\phi_\mathrm{peri}-\phi_\mathrm{b}=0$ and $\pi$, along the major axis of the bar. In this case, the orbits will be centred on the $L_1$ or $L_2$ Lagrange points. The value of $J_\phi$ at which the sign flips depends on the resonance, with the threshold at higher $J_\phi$ for larger $l/m$. It also depends on the scale length of the bar $r_\mathrm{b}$, for the reasons discussed above. We find that the CR ($l/m=0$) changes sign for some positive $J_\phi$ for all $r_\mathrm{b}\lesssim6$~kpc. This range includes all realistic lengths of the Milky Way's bar, so it is reasonable to expect that this sign change occurs in reality.

There is also an intermediate scenario when $G\Psi_1=0$ and $\ddot{\theta}_\mathrm{s}=0$. At this point resonant trapping is not possible according to perturbation theory, because the potential explored by $\theta_\mathrm{s}$ has no minima. Close to this point $\ddot{\theta}_\mathrm{s}$ is small and its potential is shallow, so resonant trapping is expected to be less important than at smaller or larger $J_\phi$. We therefore predict that the resonant lines in energy and action space (Fig.~\ref{eq:E_Lz_res}) will have gaps at certain angular momenta, where fewer stars are trapped in librating resonant orbits.

We note that these results are a simplification in the limit of a weak bar, when the resonant orbits have the same energies for all orientations. If the bar is stronger, the energies of the orbits at different pericentric orientations will bifurcate: one will have slightly larger $E$ at a given $J_\phi$. The ranges of stability of these two orbits can also overlap, with both being stable at a particular angular momentum. \citet{moreno2021} give a full treatment and calculate the stability of resonant orbits. Their Fig.~2 shows how the stability depends on the Jacobi integral $E_\mathrm{J}$, a proxy for $J_\phi$ at a given energy. However, \citet{moreno2021} is in agreement with our fundamental result: near-circular resonant orbits are stable with $\phi_\mathrm{peri}-\phi_\mathrm{b}=\pm\pi/2$, and near-radial orbits with $\phi_\mathrm{peri}-\phi_\mathrm{b}=0,\pi$. This eccentricity dependence of stable orientations was also described by \citet{fux01}.

\subsection{Radial phase space of resonant orbits}

We are now in a position to predict the observed appearance of the radial phase space $(r,v_r)$ of highly eccentric resonant orbits, taking into consideration the stable orbital orientations relative to the bar derived above. The radial velocity $v_r$ and radius $r$ of the $(l,m)$ resonant orbit are related by
\begin{align}
    v_r^2(r;L,J_\phi)=2\left[E_{(l,m)}(L,J_\phi)-\Phi_\mathrm{I}(r)\right]-\frac{L^2}{r^2},\label{eq:r_vr}
\end{align}
where $E_{(l,m)}(L,J_\phi)$ is given by equation~\ref{eq:E_Lz_res}. We calculate orbits in configuration and radial phase space in the $\theta=\pi/2$ plane for a range of $J_\phi$. \citet{belokurov_chevrons} identified substructure in $(r,v_r)$ space by isolating stars with $|L_z|<500$~kpc~km\,s$^{-1}$, so for each resonance we generate orbits in a uniform grid of $J_\phi$ in the range $-500\leq J_\phi$~[kpc~km\,s$^{-1}$]~$\leq500$. We use the equations in Appendix~\ref{section:orbits_appendix} and equation~\ref{eq:r_vr} to compute the orbits, and choose the angles of the pericentres according to the sign of $G\Psi_1$ (see Section~\ref{section:pendulum}).

\begin{figure*}
  \centering
  \includegraphics[width=\textwidth]{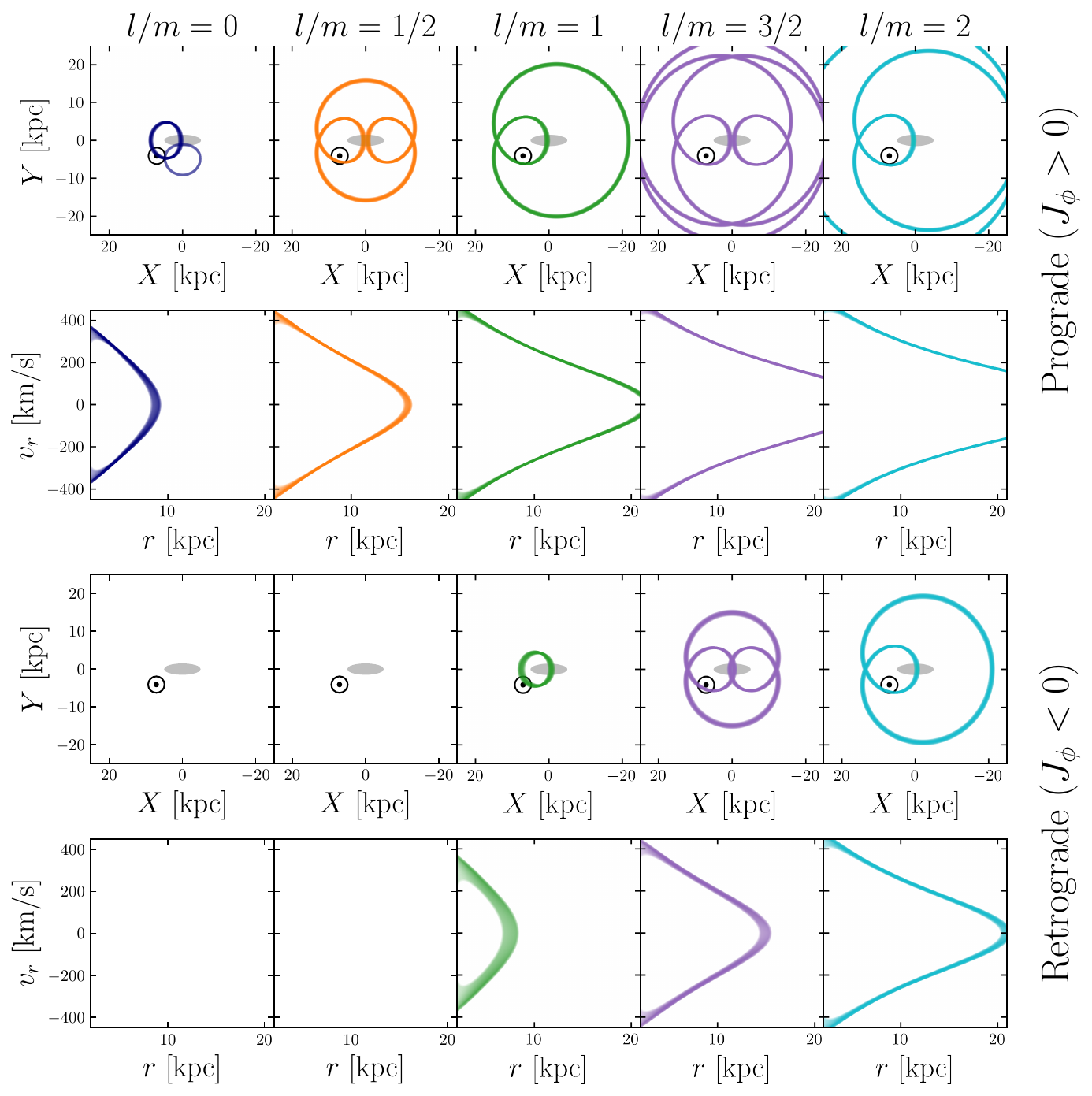}
  \caption{Resonant orbits with $|J_\phi|<500$~kpc~km\,s$^{-1}$ in configuration space $(X,Y)$ corotating with the bar (first and third rows) and in radial phase space $(r,v_r)$ (second and fourth rows). Each column shows a different resonance, and the orbits are split according to whether they are prograde (top two rows) or retrograde (bottom two rows). In the $X$-$Y$ projections, the bar (grey ellipse) lies along the $X$-axis, and the Sun is marked with $\odot$. As in Fig.~\ref{fig:cr_orbits}, the $X$-axis is flipped such that the bar rotates clockwise. These projections are therefore as viewed from the north Galactic pole.} 
  \label{fig:orbits}
\end{figure*}

We show the resonant orbits with $|J_\phi|<500$~kpc~km\,s$^{-1}$ in Fig.~\ref{fig:orbits} for $l/m\in\{0,1/2,1,3/2,2\}$, with $J_\phi>0$ (top two rows) and $J_\phi<0$ (bottom two rows). The first and third rows show the orbits in the $\theta=\pi/2$ plane, in the frame corotating with the bar (marked in grey). The $X$ and $Y$ axes are fixed to the bar's major and minor axes respectively. When there is a choice of pericentres (e.g. 0 or $\pi$), we show that which brings the orbit closest to the Sun (marked with $\odot$) in the interests of clarity. The second and fourth rows show the radial phase space, $v_r$ vs $r$.

It is important to note that motion along these orbits is in a single direction. Since the orbits are highly radial and the bar rotates clockwise in the projections shown, the motion along most of each orbit (particularly around apocentre) is anticlockwise in the corotating frame. Carefully tracing the orbits in Fig.~\ref{fig:orbits} will show that most stars will have negative $v_r$ when passing close to the Sun. This happens whenever the stable pericentre lies on the bar's major axis ($\phi_\mathrm{peri}-\phi_\mathrm{b}=0,\pi$). The only exception shown is at the CR ($l/m=0$) for $J_\phi\gtrsim400$~kpc~km\,s$^{-1}$, when $\phi_\mathrm{peri}-\phi_\mathrm{b}=\pm\pi/2$. In this case, resonant stars passing near the Sun will have $v_r>0$. This matches observations of the Hercules stream, which is likely to be a result of disc stars begin trapped in the CR \citep{Pe17, d'onghia2020}. Note also that all resonant orbits pass close to the Sun, so trapped stars should easily be visible to \textit{Gaia}.

\begin{figure}
  \centering
  \includegraphics[width=\columnwidth]{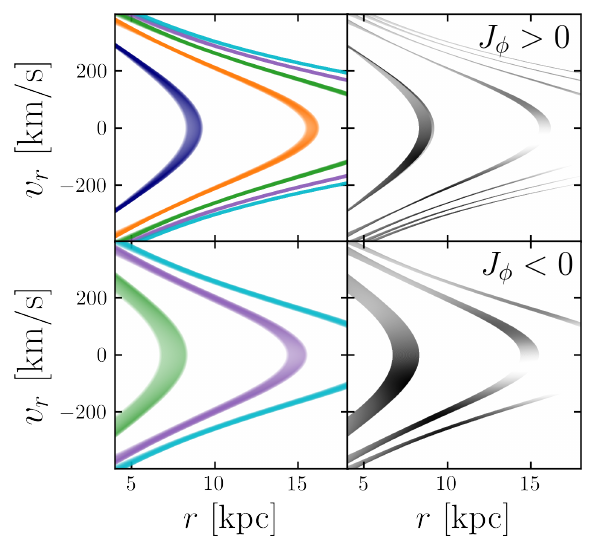}
  \caption{Radial phase space of analytic resonant orbits in the isochrone potential. The left-hand column shows each resonance colour-coded, where the colours have the same meanings as before. In the right-hand column a selection function weighting (equation~\ref{eq:selection_effect}) is applied. Darker shades correspond to higher density. On each resonance the stars are uniformly distributed in both radial angle $\theta_r$ and angular momentum $J_\phi$. We split the stars into prograde (top row) and retrograde (bottom row) subsets. As expected from the shape of the orbits (Fig.~\ref{fig:orbits}), the chevrons are asymmetric in $v_r$, with higher densities of stars at $v_r<0$ in the Solar neighbourhood.}
  \label{fig:r_vr_model}
\end{figure}

As expected from the energy distributions (Fig.~\ref{fig:E_Lz}), higher resonances of a given $J_\phi$ have apocentres at larger radii. For a given $l/m$, the retrograde resonant orbits have smaller apocentric radii, slightly less than those of the prograde $l/m-1$ orbits. By combining orbits in multiple resonances, we can therefore construct a series of nested chevrons like those observed by \citet{belokurov_chevrons}. These are shown in the left-hand column of Fig.~\ref{fig:r_vr_model}, split into prograde (top) and retrograde (bottom) orbits.

If we assume that the orbits are stable, the requirement that $\phi_\mathrm{peri}-\phi_\mathrm{b}$ take certain values severely restricts the orientations that the orbits can have. Any observational survey will naturally impose a spatial selection effect, resulting in most of its sample being located in the Solar neighbourhood. Hence, if we observe trapped stars on these highly eccentric resonant orbits, we can predict that the $(r,v_r)$ chevrons will be asymmetrical, with a greater number of stars at $v_r<0$.

We can predict the appearance of the observed radial phase space if we assume some selection function governing the number of stars in a sample as a function of Galactic position. We first take the resonant orbits with $|J_\phi|<500$~kpc~km\,s$^{-1}$ calculated above, and sample them uniformly in $\theta_r$ between 0 and $4\pi$. Since we are only considering resonant orbits with $m=2$, this gives us a steady-state distribution of particles along each resonance. We take the same number of particles from each resonance, with equal numbers having each of the stable pericentric angles (usually $0$ and $\pi$).

We calculate the heliocentric distance $D$ for each particle, and assign a weighting $S(D)$ to roughly mimic the selection function of observational data. This can be seen as the fraction of stars at a particular distance which are included in the data sample. We adopt a simple form of the weighting, namely
\begin{equation}\label{eq:selection_effect}
    S(D)\propto\mathrm{exp}\left(-\frac{D}{D_0}\right),
\end{equation}
where $D_0$ is a scale-length. We estimate $D_0$ by making the simple assumption that the true spherically averaged density of stars in a thin shell of radius $D$ centred on the Sun is independent of $D$. Then the number of stars between distances $D$ and $D+\mathrm{d}D$ is $N_\mathrm{true}(D)\mathrm{d}D\propto D^2\mathrm{d}D$. Hence the selection function $S(D)\propto N_\mathrm{obs}(D)/D^2$, where $N_\mathrm{obs}(D)\mathrm{d}D$ is the number of observed stars in the data sample between $D$ and $D+\mathrm{d}D$. An estimate for $D_0$ is obtained by plotting $\mathrm{ln}(N_\mathrm{obs}(D)/D^2)$ against $D$ and finding the approximate gradient, which is equal to $-1/D_0$. For the data sample described in Section~\ref{section:data}, we find that $D_0=1.25$~kpc is a good fit beyond distances of 2~kpc.

\begin{figure}
  \centering
  \includegraphics[width=0.9\columnwidth]{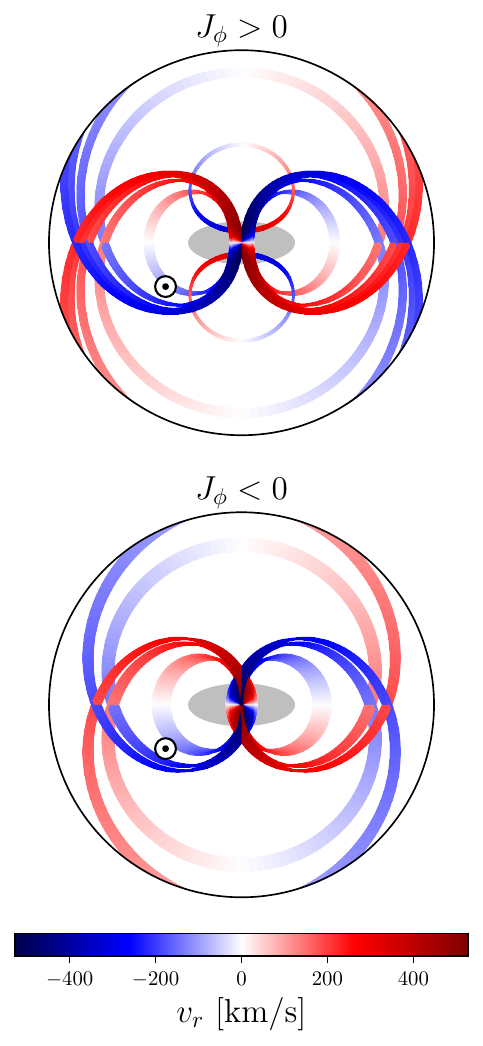}
  \caption{Resonant orbits in the Galactic plane colour-coded by radial velocity $v_r$, for prograde (top panel) and retrograde (bottom panel) orbits. Both circular panels have a radius of 18~kpc. The bar and position of the Sun are marked by the grey ellipse and the $\odot$ symbol respectively. In the Solar vicinity the resonant orbits mostly have $v_r<0$, but this changes at larger radii or different angles relative to the bar.}
  \label{fig:orbits_vr}
\end{figure}

We apply this weighting to 2D histograms of $v_r$ vs $r$ for our mock stars. These are shown in the right-hand panel of Fig.~\ref{fig:r_vr_model} for prograde (top) and retrograde (bottom) particles. As expected from the discussion above, the $v_r$ distributions in the Solar neighbourhood are asymmetric, with a greater number of stars at $v_r<0$ on almost all resonances. The only exceptions are the CR orbits with sufficiently high $J_\phi$, such that $\phi-\phi_\mathrm{peri}=\pm\pi/2$ (see Fig.~\ref{fig:GPsi_1}). However, the asymmetry switches sign at higher radius: at $r\gtrsim13$~kpc, there is instead a greater number of stars at $v_r>0$. This behaviour can be understood if we examine the radial velocity as a function of position along the resonant orbits. Fig.~\ref{fig:orbits_vr} shows the orbits calculated above in the Galactic plane, with the bar and the position of the Sun marked. The colours indicate $v_r$, with red (blue) corresponding to positive (negative) values. Near the Sun, the resonant orbits mostly have $v_r<0$, so that around $r\approx8$~kpc the chevrons are most clearly visible there. Now consider moving from the Sun away from the Galactic centre. The branches of the orbits encountered at larger radii at the Sun's azimuth instead have $v_r>0$. This results in the asymmetry changing sign at larger radii, as is seen in Fig.~\ref{fig:r_vr_model}.

We have used our analytical model to predict the appearance of the radial phase space of highly eccentric resonant orbits in the Milky Way. In the next section we compare these predictions to the results of a test particle simulation of bar-generated chevrons.

\section{Test particle simulations}
\label{section:simulations}

\subsection{Setup}

To test the qualitative predictions of our analytical model we run a test particle simulation in a realistic Milky Way potential with a bar. This includes extra effects that may be important in reality, such as 3D orbits and a stronger bar. We follow a similar procedure to that described in \citet{dillamore2023}, except that we use a different potential and a decelerating bar. This is expected to increase the number of stars trapped in resonances \citep{chiba2021}, so that they will be more prominent in phase space. The setup of our simulations is summarised below and described in more detail in Appendix~\ref{section:sim_appendix}.

We use the Milky Way potential fitted by \citet{sormani2022} to the Milky Way model of \citet{Po17}. This consists of an axisymmetric disc and three bar components. There is in addition a flattened axisymmetric dark halo and a central mass concentration (representing a nuclear stellar disc or cluster). The Solar radius and corresponding circular velocity are 8.2~kpc and 238~km\,s$^{-1}$ respectively, as used in our analytical model. As in Section~\ref{section:model} we place the Sun at an angle of $-30^\circ$ relative to the bar.

A realistic decelerating bar should increase in length as it slows, roughly scaling with the corotation radius \citep[e.g.][]{athanassoula1992}. We therefore modify the potential by changing the scale-length of its non-axisymmetric multipole components (those representing the bar). Specifically, the scale is inversely proportional to the bar's pattern speed $\Omegab$, and matches that of the \citet{sormani2022} potential when $\Omegab=39$~km\,s$^{-1}$kpc$^{-1}$ (the pattern speed of the \citet{Po17} model). We leave the amplitude of these components unchanged during this scaling.

Between times $t=0$ and $t=2$~Gyr of the simulation we smoothly increase the strength of the non-axisymmetric components of the potential from zero to their full values. The axisymmetric components are kept constant so that the total mass is conserved. We use the same prescription for this smooth increase as \citet{dehnen2000} and \citet{dillamore2023}.

We set the initial pattern speed to 80~km\,s$^{-1}$kpc$^{-1}$ and smoothly decrease it after the bar is fully grown in amplitude. Following \citet{chiba2021} we quantify the slowing rate with the dimensionless deceleration parameter, $\eta\equiv-\dot{\Omega}_\mathrm{b}/\Omegab^2$. This is held constant at $\eta=0.003$ after $t=3$~Gyr, within the range of best-fitting values calculated by \citet{chiba2021}. The full equations governing the evolution of the bar strength and $\Omegab(t)$ are given in Appendix~\ref{section:sim_appendix}.

We initialise a distribution of stars using a steady-state distribution function (DF) in the initially axisymmetric potential. The DF $f=f_\mathrm{halo}+f_\mathrm{disc}$ consists of two components: a non-rotating spheroidal component $f_\mathrm{halo}$ and a flattened rotating component $f_\mathrm{disc}$.

The halo DF $f_\mathrm{halo}$ expressed as a function of the actions is derived from the spherical density profile
\begin{equation}
    \rho_\mathrm{halo}\propto\left(\frac{r}{r_\mathrm{s}}\right)^{-\gamma}\left[1+\left(\frac{r}{r_\mathrm{s}}\right)^{\gamma-\beta}\right]
\end{equation}
using an isotropic \texttt{QuasiSpherical} distribution function \citep{Je17,agama}. We set the inner and outer power law slopes to $\gamma=2.3$ and $\beta=4.6$ respectively \citep{deason2011_halo}, and the break radius to $r_\mathrm{s}=25$~kpc \citep{Wa09,deason2011_halo,faccioli2014_halo,pila-diez2015}. 

The flattened rotating component is generated using the exponential DF \citep{agama},
\begin{align}
\begin{split}
    f_\mathrm{disc}(\textbf{\textit{J}})&\propto\tilde{J^3}\;\mathrm{exp}\left(-\frac{\tilde{J}}{J_{\phi,0}}\right)\mathrm{exp}\left(-\frac{\tilde{J}J_r}{J_{r,0}^2}\right)\mathrm{exp}\left(-\frac{\tilde{J}J_z}{J_{z,0}^2}\right)\\&\times
    \begin{cases}
      1 & J_\phi>0 \\
      \mathrm{exp}\left(\frac{\tilde{J}J_\phi}{J_{r,0}^2}\right) & J_\phi<0,
   \end{cases}\\
   \tilde{J}&\equiv|J_\phi|+J_r+0.25J_z,
\end{split}
\end{align}
where $\{J_{r,0}$, $J_{z,0}$, $J_{\phi,0}\}=\{1100,200,400\}$~kpc$\,$km$\,$s$^{-1}$. Although this is nominally a thick disc-like component, the high radial action scale $J_{r,0}$ gives us a large number of stars on highly eccentric halo-like orbits that we are interested in studying. We use a total of $\approx2\times10^6$ particles, roughly equally split between the disc and halo components.

We integrate the orbits of the particles from the initial DF $f$ in the evolving barred potential. We run the simulation for total time $t_\mathrm{f}\approx6.6$~Gyr, at which point the pattern speed is $\Omegab=35$~km\,s$^{-1}$kpc$^{-1}$ (as used in our analytical model).

\subsection{Results}
\begin{figure}
  \centering
  \includegraphics[width=\columnwidth]{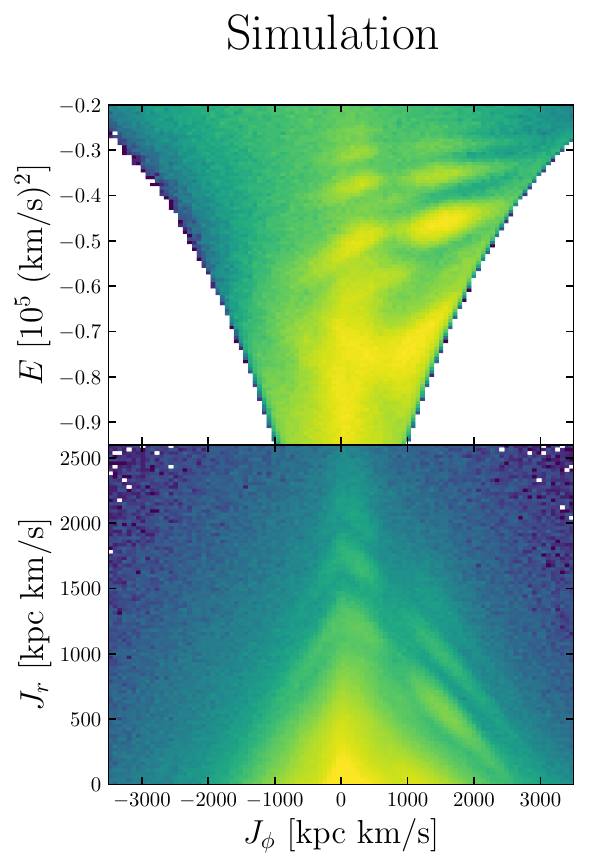}
  \caption{Energy $E$ (top panel) and radial action $J_r$ (bottom panel) vs $J_\phi$ in the simulation. In both cases the resonances are visible as overdense ridges. These have a very similar form to those shown for the analytical isochrone model in Fig.~\ref{fig:E_Lz}.}
  \label{fig:E_Lz_sim}
\end{figure}

In Fig.~\ref{fig:E_Lz_sim} we show the energy $E$ and radial action $J_r$ vs the $z$-component angular momentum $L_z=J_\phi$ at the final snapshot of the simulation. The radial action is calculated in the axisymmetrised (i.e. azimuthally averaged) \citet{Po17} potential using \textsc{agama} \citep{agama}. This is analogous to Fig.~\ref{fig:E_Lz}. Indeed, in both panels there is a good match with the analytic isochrone model. In particular, in $E$ vs $J_\phi$ space the ridges have roughly constant energy for $J_\phi>0$ but tend to lower energies as $J_\phi$ decreases towards and below zero. Similarly, in the lower panel the ridges are steeper at $J_\phi<0$ than their prograde counterparts. Our isochrone model therefore successfully predicts the qualitative appearance of bar resonances in integral of motion space, including the presence of retrograde resonances. Note that there are gaps in the resonant ridges at $J_\phi\sim800$~kpc~km\,s$^{-1}$. These seem to confirm the prediction made by the resonant perturbation theory that trapping is not possible at certain values of $J_\phi$. If this is the case, these gaps should also coincide with the points at which the stable $m=2$ resonant orbits change orientation by $\pi/2$. This is confirmed in Section~\ref{section:analysis_iom_space}.

\begin{figure}
  \centering
  \includegraphics[width=\columnwidth]{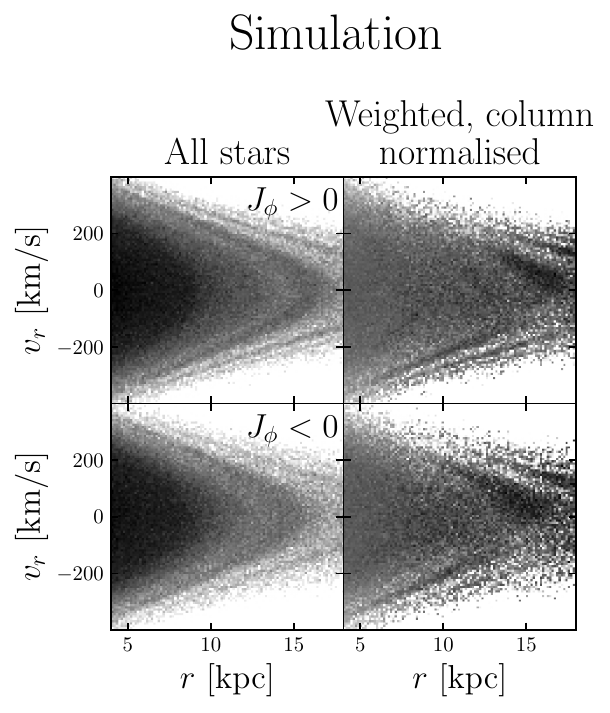}
  \caption{Radial phase space of the simulation for prograde (top row) and retrograde (bottom row) stars with $|J_\phi|<500$~kpc~km\,s$^{-1}$. The left column shows the unweighted distribution, while in the right-hand column we apply the selection effect weighting defined by equation~\ref{eq:selection_effect}. We also column-normalise the weighted histogram, such that the total count along each column of $v_r$ bins is the same. The weighting results in the resonant chevrons becoming asymmetric in $v_r$.}
  \label{fig:r_vr_sim}
\end{figure}

We now turn to the radial phase space. In Fig.~\ref{fig:r_vr_sim} we show $v_r$ vs $r$ for stars with $|J_\phi|<500$~kpc~km\,s$^{-1}$. The top and bottom rows show prograde and retrograde stars respectively. The unweighted simulation data is shown in the left-hand column. As shown by \citet{dillamore2023} and predicted by our model in Section~\ref{section:model}, the resonances manifest as chevron-shaped overdensities. While our simulations in \citet{dillamore2023} only produced one clear overdensity, our setup here is able to produce multiple nested chevrons. This is due to a combination of the slowing bar and the flattened rotating component of the initial distribution function. In the prograde case (top-left panel) at least three or four different resonant chevrons are visible, which matches or exceeds the number observed in the Milky Way by \citet{belokurov_chevrons}. Multiple resonant chevrons are also visible for retrograde orbits (bottom-left panel). Since our isochrone potential is a good match at most radii to the \citet{Po17} potential used in the simulations (see Fig.~\ref{fig:rot_curve}), it is possible to associate each chevron with a specific resonance by comparing with the left-hand column of Fig.~\ref{fig:r_vr_model}. For example, the prograde chevron peaking at $r\approx16$~kpc is due to the OLR ($l/m=1/2$), and its retrograde counterpart peaking at $r\approx15$~kpc is due to the $l/m=3/2$ resonance.

In the right-hand column we apply the weighting defined by equation~\ref{eq:selection_effect}. The effect is similar to that predicted by the model in the right-hand column of Fig.~\ref{fig:r_vr_model}: in the Solar neighbourhood ($r\sim8$~kpc), the chevrons are only clearly visible at $v_r<0$. The reverse is true for $r>13$~kpc, and the peak of the OLR chevron mentioned above is most evident at $v_r>0$.

We therefore see that asymmetry in $v_r$ is a natural consequence of observing chevrons generated by bar resonances in some volume-limited region. The above analysis also provides us with a test: if the chevrons observed in the Milky Way are formed by resonances, the $v_r$ asymmetry should change as a function of position. This result is not predicted by the explanation of \citet{donlon2023} for the observed asymmetry. We now proceed to study data from \textit{Gaia} in order to test this prediction.

\section{Data}\label{section:data}

\begin{figure}
  \centering
  \includegraphics[width=\columnwidth]{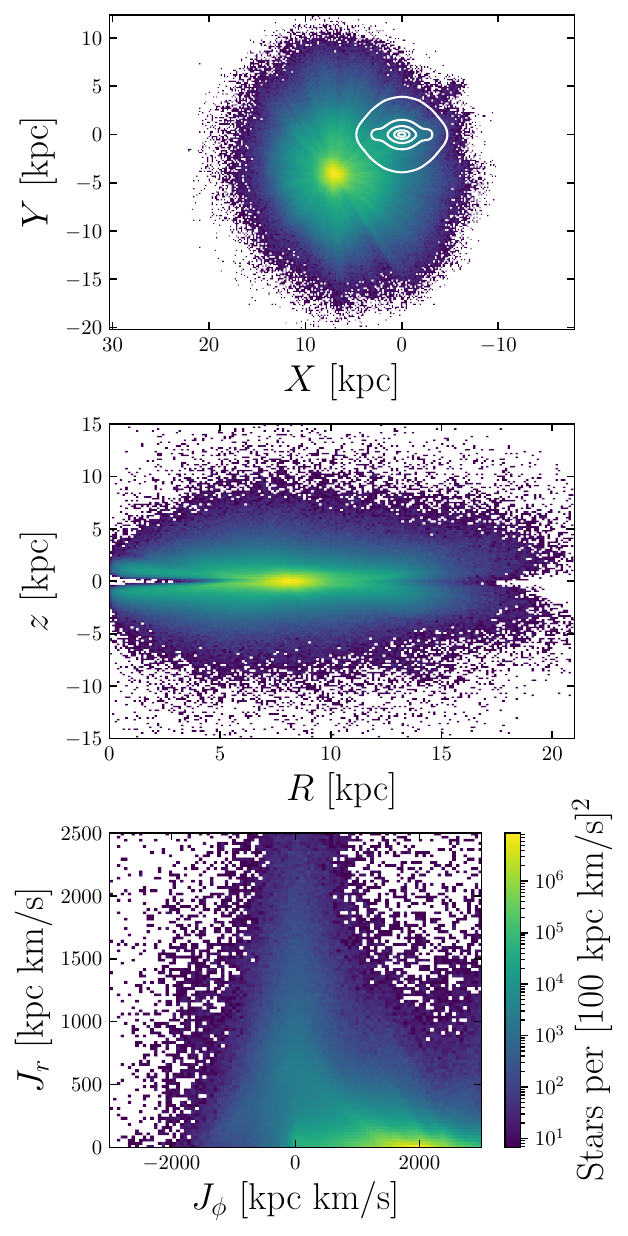}
  \caption{Spatial distributions of the data sample. \textbf{Top panel:} Face-on projection of the Galactic plane in Galactocentric Cartesian coordinates. The bar lies along the $X$-axis and is indicated by the white contours. \textbf{Middle panel:} Height above the Galactic plane $z$ versus cylindrical radius $R$. \textbf{Bottom panel:} radial action $J_r$ vs azimuthal action $J_\phi$.}
  \label{fig:data_dists}
\end{figure}

\begin{figure}
  \centering
  \includegraphics[width=\columnwidth]{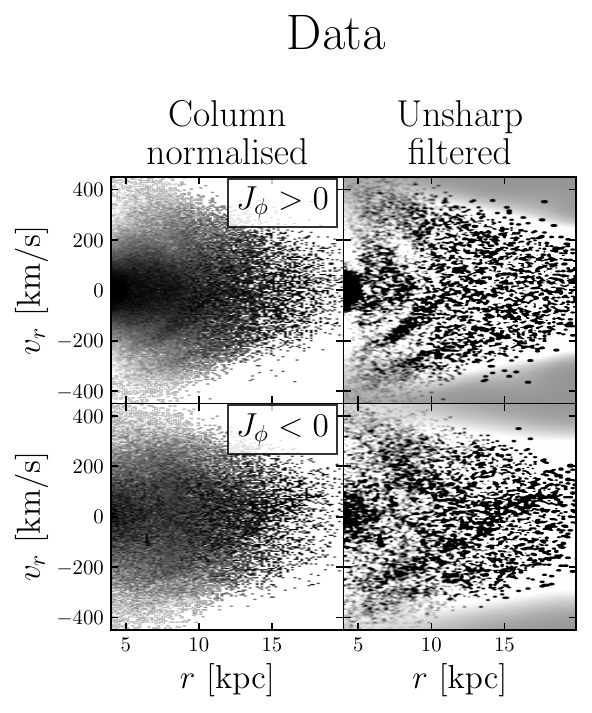}
  \caption{Radial phase space $(r, v_r)$ of the observed data with $|J_\phi|<500$~kpc~km\,s$^{-1}$. The top (bottom) row shows prograde (retrograde) stars. In the left-hand column the data is column-normalised, such that the total count across all $v_r$ bins in the histogram is independent of $r$. In the right-hand column a smooth background has been subtracted. This background was calculated by convolving the column-normalised histogram with Gaussians of standard deviation $\sigma=8$ ($J_\phi>0$) and 16 ($J_\phi<0$) pixels. This unsharp masking reveals the chevrons discovered by \citet{belokurov_chevrons}.}
  \label{fig:r_vr_data}
\end{figure}

We use positions, proper motions and line-of-sight velocities from the third data release (DR3) of \textit{Gaia} \citep{gaia,gaia_dr3}. The line-of-sight velocity measurements are provided by the Radial Velocity Spectrometer \citep[RVS;][]{gaia_rvs} We supplement these with updated estimates of parallax calculated from \textit{Gaia} DR3 XP spectra by \citet{zhang2023}. These parallax estimates provide us with distance measurements for stars up to larger Galactic radii than those of \citet{bailer-jones2021}, as were used by e.g. \citet{belokurov_chevrons}, which allows us to study a wider range of the observed radial phase space. The revised distance measurements are $D=1/\varpi$, where $\varpi$ are the parallax estimates provided by \citet{zhang2023}.

We apply the following cuts to clean the data sample. Following the recommendation of \citet{zhang2023}, we use only stars from their catalogue with \texttt{quality\_flags}~$<8$. We select stars with distance uncertainties of less than 10\% (i.e. $\varpi/\sigma_\varpi>10$, where $\sigma_\varpi$ is the parallax uncertainty), and distances $D<15$~kpc. We also remove sources with an angular separation of less than $1.5^\circ$ from known globular clusters within 5~kpc of the Sun.


Our final sample consists of $\approx22$ million stars with 6D phase space positions and stellar parameter estimates by \citet{zhang2023}. In order to directly compare to our model and simulations, we transform to a left-handed Galactocentric coordinate system ($X$, $Y$, $z$) in which the Sun is located in the Galactic plane at a distance $r_0=8.2$~kpc from the Galactic centre. The local standard of rest (LSR) is assumed to move on a circular orbit with velocity $v_0=238$~km\,s$^{-1}$ \citep{bland-hawthorn2016}, and the Sun's motion relative to the LSR is $(U,V,W)_\odot=(11.1,12.24,7.25)$~km\,s$^{-1}$ \citep{schonrich2010}. These match the parameters used by \citet{Po17}. We align the $X$-axis (and $\phi=0$ plane) of this coordinate system with the major axis of the bar, and place the Sun at an azimuth of $\phi_\odot=-30^\circ$. These parameters are identical to those used in the isochrone model and simulations in Sections~\ref{section:model} and \ref{section:simulations}. In this left-handed coordinate system the bar rotates with positive angular frequency.

The distribution of the final sample is shown in Fig.~\ref{fig:data_dists}. The top panel shows the Galactic plane face-on, with the Sun located at $(X,Y)=(7.1,-4.1)$~kpc. Note that the $X$-axis is flipped such that the disc and bar rotate clockwise in this view. The white contours show the projected density of the \citet{sormani2022} bar model for comparison. The middle panel shows height above the plane $z$ against cylindrical radius $R$. Our sample extends up to and beyond $R\approx18$~kpc and $|z|\approx5$~kpc, well into the halo. It also covers a wide range of Galactic azimuth $\phi$. The bottom panel shows the radial vs azimuthal action space $(J_\phi,J_r)$. While much of the sample is at low $J_r$, it also contains many stars with eccentric and retrograde orbits, as required for our study.

\subsection{Radial phase space}

In Fig.~\ref{fig:r_vr_data} we show the density in radial phase space $(r,v_r)$ for the subset of the data with angular momentum $|J_\phi|<500$~kpc~km\,s$^{-1}$. In the left-hand column the histograms are column-normalised, such that each $r$ bin has the same total count across all $v_r$ bins. The chevrons observed by \citet{belokurov_chevrons} are already visible as diagonal overdense striations, particularly at $J_\phi>0$ and $v_r<0$. Following \citet{belokurov_chevrons}, in the right-hand column we apply unsharp masking to reveal this finer structure. We first convolve the histograms with Gaussians with standard deviations of 8 pixels (prograde) and 16 pixels (retrograde) along $r$ and $v_r$, then subtract these from the original histograms. This process reveals the chevrons more clearly, with at least two being visible for both prograde and retrograde stars.

While the shape of the chevrons resembles those produced by highly radial merger events \citep{dong-paez2022,davies2023_ironing,davies2023_bar,donlon2023}, there is also a striking asymmetry in $v_r$, particularly at $J_\phi>0$. While the chevrons can clearly can be seen at $v_r<0$, they are virtually invisible at $v_r>0$. The presence of $v_r$ asymmetry was previously noted by \citet{belokurov_chevrons} and used by \citet{donlon2023} as evidence that the prograde chevrons are dynamically young. This is because chevrons from a recent merger reach a maximum radius at a positive value of $v_r$ \citep{dong-paez2022}, so are asymmetrical in $v_r$.

However, this is not what is observed in the Milky Way. While the chevrons in the top-right panel of Fig.~\ref{fig:r_vr_data} are indeed asymmetric, they are barely visible at $v_r>0$. This is not what is expected from asymmetric dynamically young chevrons, which are similarly overdense at both positive and negative $v_r$ \citep[e.g. see Fig.~2 in][]{donlon2023}. The observed asymmetry appears to be of a different nature.

However, the observed radial phase space does resemble that predicted by our bar resonance model in Figs.~\ref{fig:r_vr_model} and \ref{fig:r_vr_sim}. As seen in the weighted distributions from the simulation (Fig.~\ref{fig:r_vr_sim}, right-hand column), little or no substructure is observed at $v_r>0$ in the Solar neighbourhood ($r\approx4-12$~kpc). This is also consistent with the analytic isochrone model, in which we predicted that around the Solar radius the chevrons would be more prominent at $v_r<0$. However, the most prominent retrograde chevron is more symmetric, as previously noted by \citet{donlon2023}.

On first inspection the observed chevrons match the predictions for resonance-generated chevrons, at least for prograde stars. In the next section we directly compare the analytic predictions, simulation and observations in more detail.

\section{Analysis}
\label{section:analysis}

\begin{figure*}
  \centering
  \includegraphics[width=\textwidth]{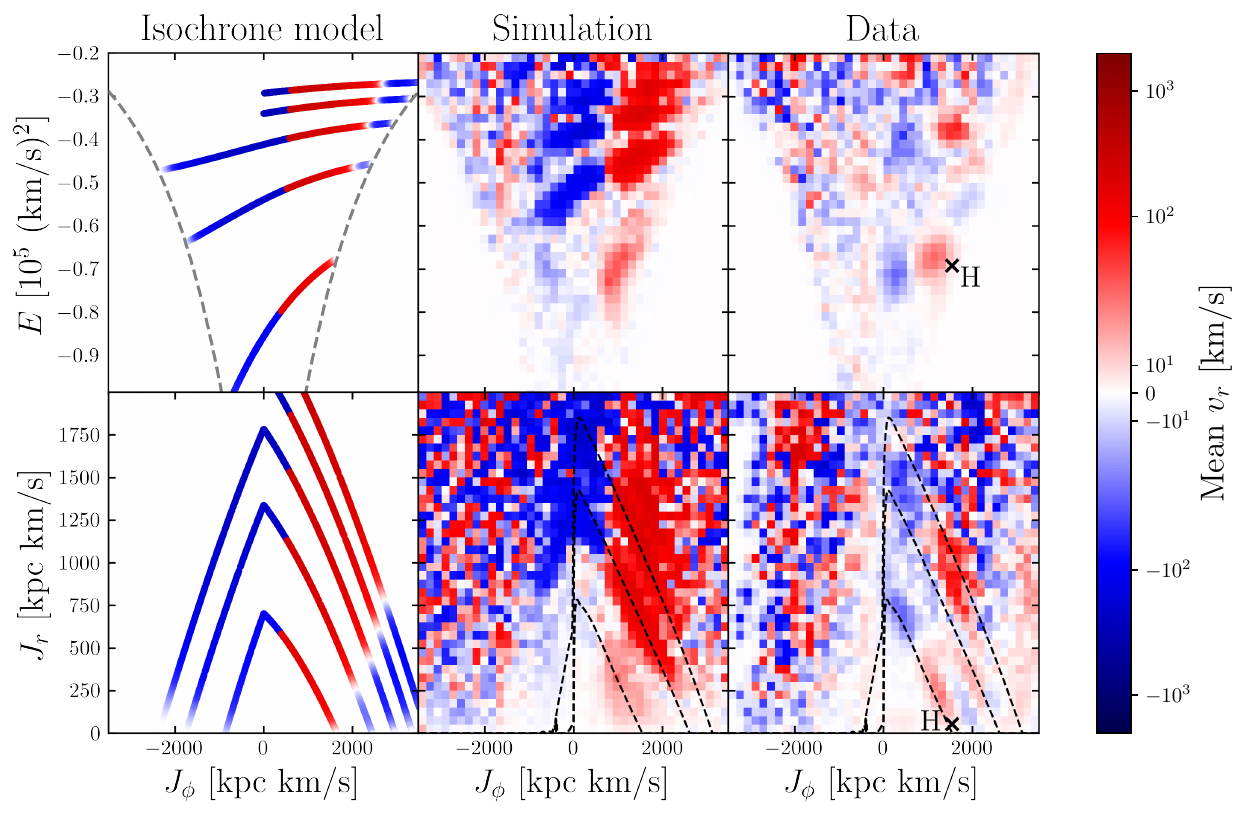}
  \caption{Energy (top row) and radial action (bottom row) versus angular momentum $J_\phi$, colour coded by mean radial velocity $v_r$. The left-hand column shows the analytical isochrone model, the middle the simulation and the right the data. The model and simulation are both weighted according to equation~\ref{eq:selection_effect} to mimic the selection effects of the data. The black dashed lines in the action space panels of the simulation and data mark the CR, OLR and 1:1 resonances calculated for a pattern speed of $\Omegab=35$~km\,s$^{-1}$kpc$^{-1}$. The crosses labelled `H' denote the approximate location of the Hercules stream.} 
  \label{fig:E_Lz_mean_vr}
\end{figure*}

\begin{figure}
  \centering
  \includegraphics[width=0.95\columnwidth]{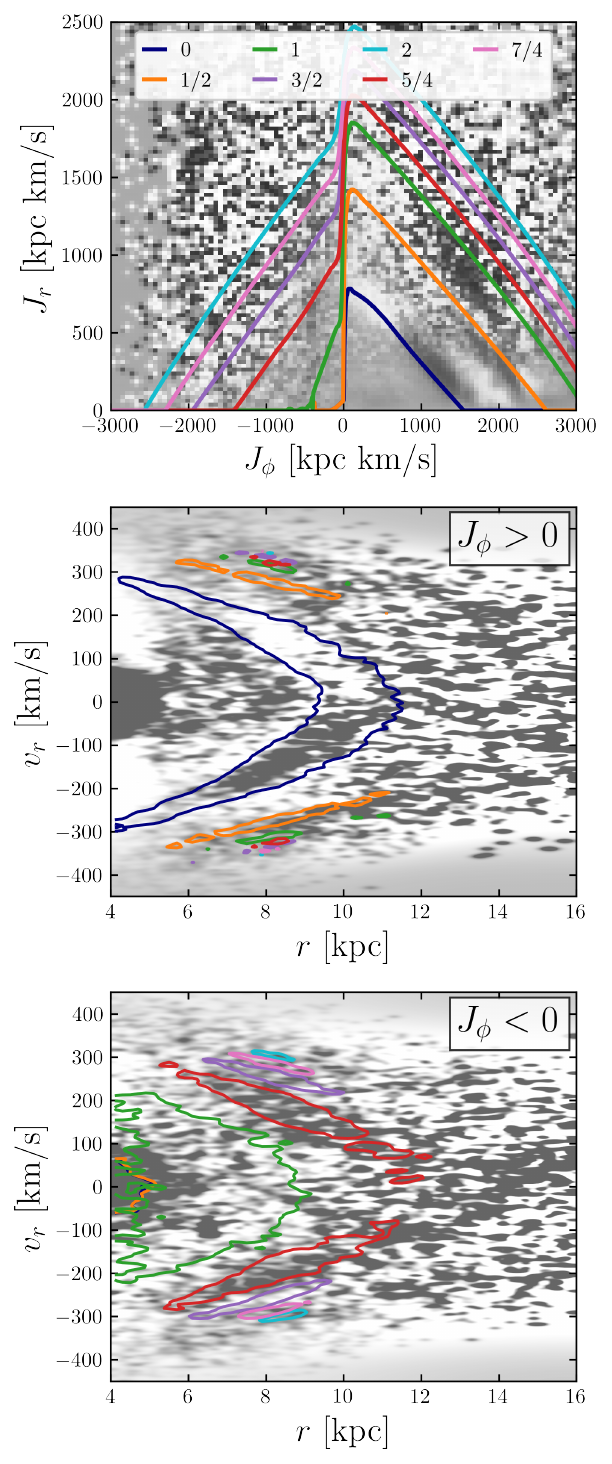}
  \caption{\textbf{Top panel:} a repeat of part of the bottom-right panel of Fig.~\ref{fig:E_Lz_mean_vr}, showing average radial velocity as a function of action. Here black (white) pixels correspond to positive (negative) mean $v_r$. The coloured lines indicate resonances with a bar rotating at pattern speed $\Omegab=35$~km\,s$^{-1}$kpc$^{-1}$. The numbers in the legend indicate the resonant ratio $l/m$. \textbf{Middle and bottom panels:} radial phase space of the data as seen in Fig.~\ref{fig:r_vr_data}, for prograde and retrograde stars respectively. Density contours of the selected stars near resonances are plotted on top. Some of these correspond very closely with the most prominent chevrons.}
  \label{fig:data_resonances}
\end{figure}

\begin{figure}
  \centering
  \includegraphics[width=\columnwidth]{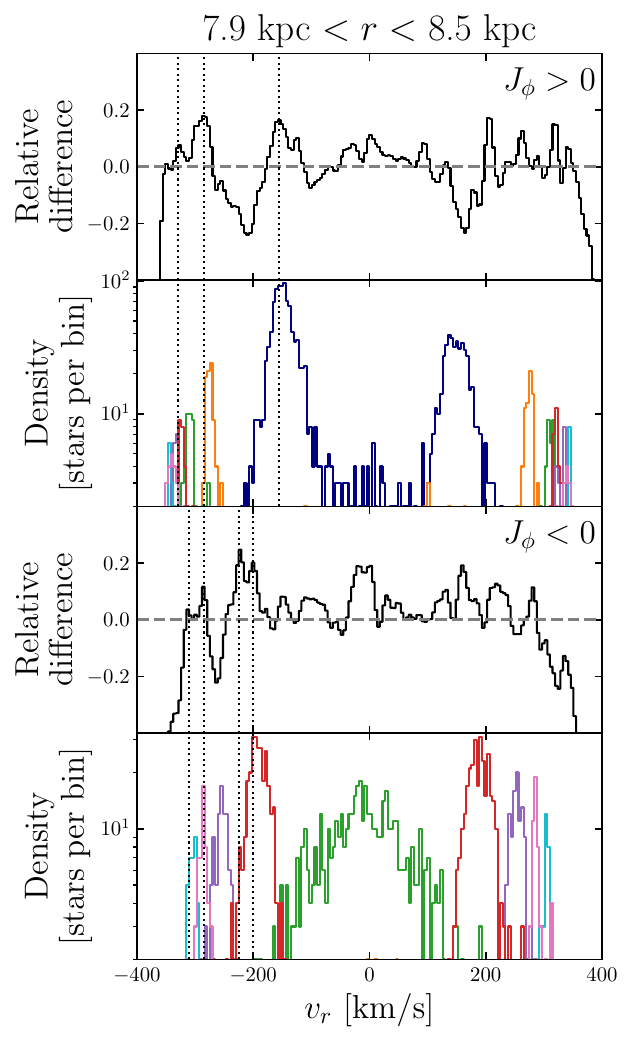}
  \caption{Distributions of $v_r$ in the radial slice 7.9~kpc~$<r<$~8.5~kpc and $|J_\phi|<500$~kpc~km\,s$^{-1}$. The top two and bottom two panels show prograde and retrograde stars respectively. The first and third panels show the relative difference between the distribution of all stars in this slice and a smooth background (i.e. convolved with a Gaussian). The coloured histograms in the second and bottom panels show those selected close to the resonances. The colours have the same meanings as in Fig.~\ref{fig:data_resonances}. Some of the strongest peaks are marked with black dotted lines.}
  \label{fig:vr_slice}
\end{figure}

\begin{figure*}
  \centering
  \includegraphics[width=\textwidth]{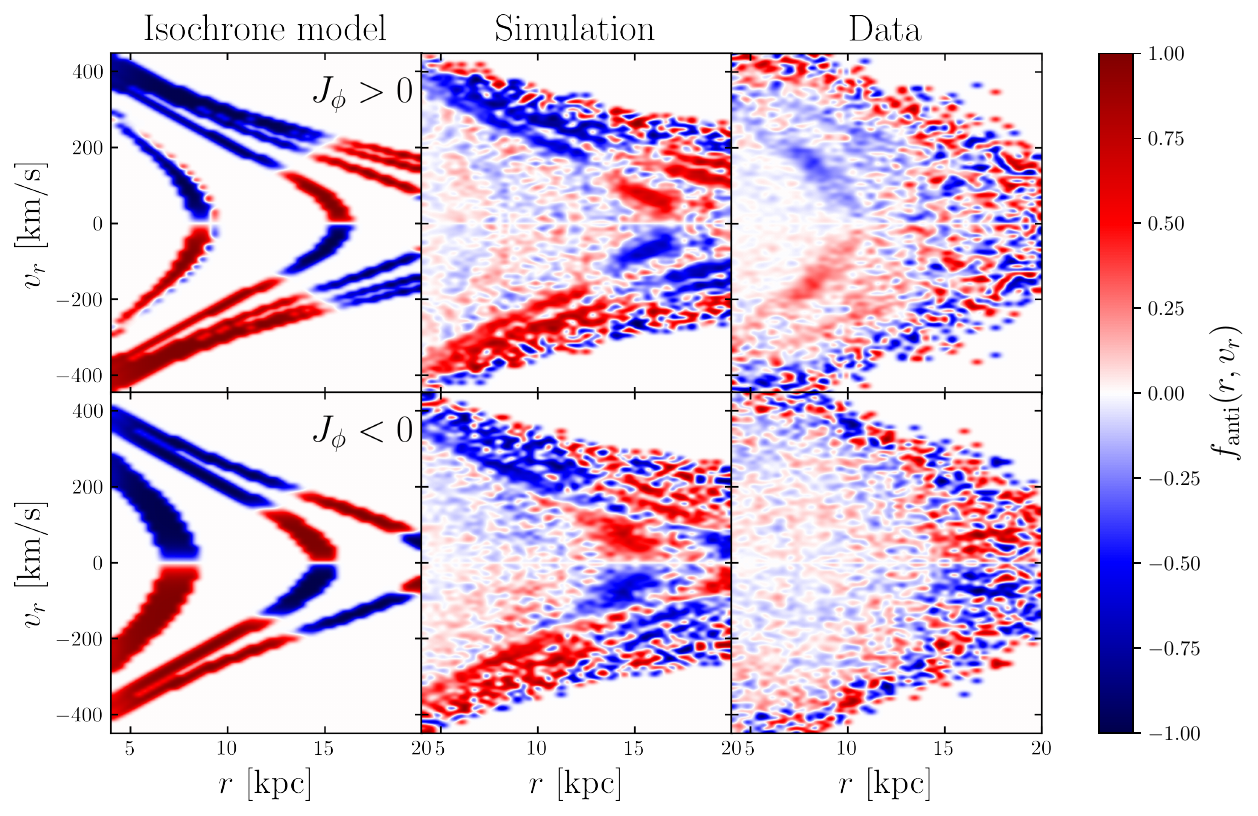}
  \caption{Antisymmetric component of the radial phase space distributions, as defined by equation~\ref{eq:r_vr_antisym}. The left-hand, middle and right-hand columns show the analytic model, the simulation and the data respectively. The top (bottom) row shows prograde (retrograde) stars. Red (blue) indicates an excess (deficit) of stars compared to the opposite value of $v_r$.}
  \label{fig:r_vr_antisym}
\end{figure*}

\begin{figure*}
  \centering
  \includegraphics[width=\textwidth]{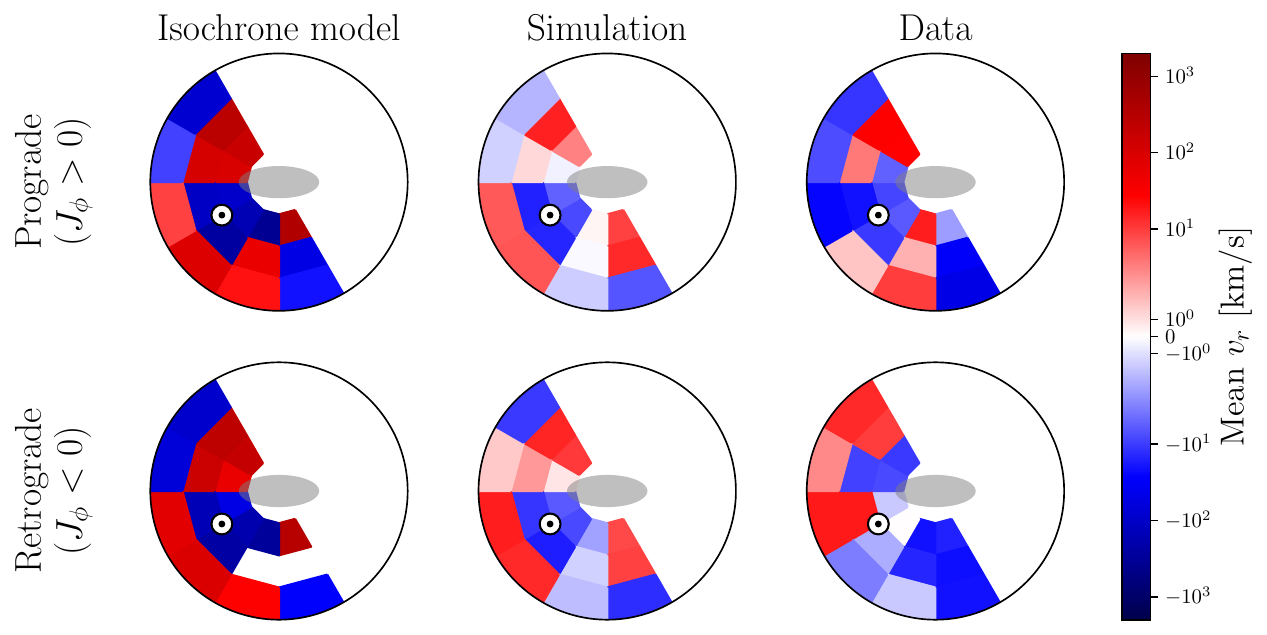}
  \caption{Mean $v_r$ as a function of Galactic position for the isochrone model (left-hand column), the simulation (middle column) and the data (right-hand column). Each circle has a radius of 18~kpc, and the $\odot$ symbol marks the position of the Sun. The top and bottom rows show prograde (retrograde) stars. Red (blue) indicates a positive (negative) mean $v_r$.}
  \label{fig:mean_vr_position}
\end{figure*}

\begin{figure}
  \centering
  \includegraphics[width=\columnwidth]{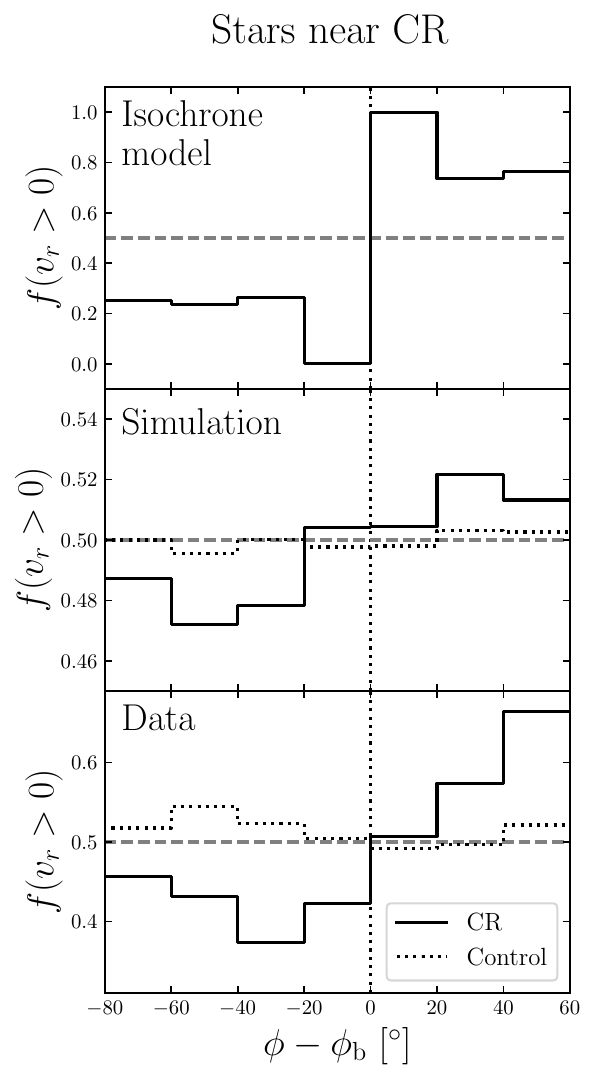}
  \caption{Fraction of stars near the corotation resonance with $v_r$ as a function of azimuth. The dotted distributions in the lower two panels show control samples (i.e. stars not near CR). For the data a bar pattern speed of $\Omegab=35$~km\,s$^{-1}$kpc$^{-1}$ is assumed. The vertical black dotted line marks the major axis of the bar, and the horizontal grey dashed line indicates an equal number of stars with positive and negative $v_r$.}
  \label{fig:vr_asymmetry_phi}
\end{figure}

We have so far presented the radial phase space of the analytic isochrone model, the simulations and the data. We have shown that asymmetry in $v_r$ is a characteristic feature of resonant chevrons, and that this is also seen in the data. We now use this asymmetry to more closely investigate the chevrons in various projections of phase space. We also associate the chevrons with specific resonances.

\subsection{Integral of motion space}
\label{section:analysis_iom_space}

We calculate the energy $E$ and actions $J_\phi$ and $J_r$ of stars in the data sample using \textsc{agama}. As in the simulations, we use the axisymmetrised \citet{sormani2020} potential. In Fig.~\ref{fig:E_Lz_mean_vr} we show energy $E$ (top row) and radial action $J_r$ (bottom row) vs $J_\phi$ for the isochrone model, the simulation and the data. We have coloured each pixel with the mean radial velocity $v_r$ of all stars in that pixel. In the case of the isochrone model and simulation, this is the weighted mean where equation \ref{eq:selection_effect} has been used as the weighting. We also calculate the actions $J_r$ of the CR, OLR and 1:1 resonant orbits in the Galactic plane by numerically solving equation~\ref{eq:resonances} at a range of $J_\phi$ values. As in the isochrone model and simulation, we set the pattern speed to $\Omegab=35$~km\,s$^{-1}$kpc$^{-1}$. The loci of these resonances are marked by black dashed lines. We also calculate the approximate actions and energies of the Hercules stream assuming velocities of $(v_r,v_\phi)=(30, 188)$~km\,s$^{-1}$ \citep{antoja2008}, and indicate these with the cross labelled with `H' in each data panel.

Qualitatively the isochrone model and simulation show the same behaviour. As $J_\phi$ increases along each resonance, the mean $v_r$ changes sign from negative (at $J_\phi\lesssim800$~kpc~km\,s$^{-1}$) to positive (up to $J_\phi\approx2000-3000$~kpc~km\,s$^{-1}$) then back to negative at higher $J_\phi$. The change in sign at $J_\phi\approx800$~kpc~km\,s$^{-1}$ can be understood from the reasoning in Section~\ref{section:pendulum}. We showed that the stable orientation of the orbit's pericentre relative to the bar changes at roughly this value of $J_\phi$. Due to the shapes of the orbits this leads to the particles passing through the Solar neighbourhood having $v_r<0$ for smaller $J_\phi$ and $v_r>0$ for larger $J_\phi$ (e.g. see Fig.~\ref{fig:orbits_vr} for orbits with small $|J_\phi|$). The sign changes of $v_r$ also approximately coincide with the gaps in the resonant ridges seen in Fig.~\ref{fig:E_Lz_sim}, where the trapping potential is close to flat (Equation~\ref{eq:pendulum}). The change back to a negative mean $v_r$ occurs for the OLR and higher resonances, and is due to the shape of the orbit changing as it becomes more circular. As $J_\phi$ increases, the loops in the orbits (e.g. see Fig.~\ref{fig:orbits}, top row) become smaller, which results in the closest stars to the Sun having $v_r<0$. The fact that disc-like orbits at the OLR have a negative average $v_r$ has been well documented \citep[e.g.][]{trick2021,chiba2021}.

A characteristic signature of prograde resonances in action space ($J_r$ vs $J_\phi$) is therefore a series of diagonal bands with negative gradient, along which the sign of the mean $v_r$ changes from negative ($J_\phi$ small) to positive ($J_\phi$ larger). Qualitatively this is also seen in the data. The three resonances marked coincide with such bands, where the mean $v_r$ changes from negative to positive as $J_\phi$ increases and $J_r$ decreases. The value of $J_\phi$ at which this transition occurs is similar to the simulation for the CR and 1:1 resonances (lowest and highest dashed lines), while for the OLR (middle line) this transition occurs at a higher value of $J_\phi$. These bands can also be seen in the top-right panel in the space of $E$ vs $J_\phi$.

The Hercules stream is also very close to the CR and the associated diagonal band in action space. This is unsurprising, since the Hercules stream has previously been found to be consistent with trapping in the CR at a pattern speed of $\Omegab=35$~km\,s$^{-1}$kpc$^{-1}$. However, this band can be traced to much lower $J_\phi$ and higher $J_r$ than has been previously noted. The mean $v_r$ also transitions from positive in the vicinity of the Hercules stream to negative at lower $J_\phi$. For this pattern speed the CR lies along the whole length of this band. We can therefore directly connect the Hercules stream to much more eccentric halo-like orbits on the same resonance, but with negative mean $v_r$ in the Solar neighbourhood.

\subsection{Identification of resonances in radial phase space}
\label{section:data_resonances}

We have demonstrated that for a pattern speed of $\Omegab=35$~km\,s$^{-1}$kpc$^{-1}$, resonances with the bar coincide with both the Hercules stream and the diagonal stripes in the action space of the data. We now return to the radial phase space of the data and compare the chevrons with the distributions of stars near the resonances. The top panel of Fig.~\ref{fig:data_resonances} again shows the action space of the data colour-coded by mean $v_r$, as in the bottom-right panel of Fig.~\ref{fig:E_Lz_mean_vr}. We again show the approximate tracks of periodic resonant orbits in the Galactic plane, calculated in the axisymmetric potential for $\Omegab=35$~km\,s$^{-1}$kpc$^{-1}$. For each resonance we select stars with $|J_\phi|<500$~kpc~km\,s$^{-1}$ for which $|J_r-J_{r,\mathrm{res}}(J_\phi)|<50$~kpc~km\,s$^{-1}$, where $J_{r,\mathrm{res}}(J_\phi)$ is the radial action of the resonance at $J_\phi$. We show contours of the distributions of these stars in radial phase space for $J_\phi>0$ (middle panel) and $J_\phi<0$ (bottom panel).

In Fig.~\ref{fig:vr_slice} we show slices of the radial phase space in the vicinity of the Sun, between $r=7.9$ and 8.5~kpc. The top two (bottom two) panels show prograde (retrograde) stars. In the first and third panels we show the relative difference between the distributions of $v_r$ and a smooth background (i.e. difference divided by background). As in Figs.~\ref{fig:r_vr_data} and \ref{fig:data_resonances} this background is calculated by convolving the distribution with a Gaussian. The second and bottom panels show the $v_r$ distributions of stars selected to be near the resonances, similarly to in Fig.~\ref{fig:data_resonances}.

At $J_\phi>0$, Fig.~\ref{fig:data_resonances} shows that the stars close to the corotation resonance (blue) coincide remarkably closely with the innermost chevron \citep[called Chevron 1 by][]{belokurov_chevrons,dillamore2023}. The outer chevron (Chevron 3) also coincides with stars near the OLR (orange). This is confirmed in the top two panels of Fig.~\ref{fig:vr_slice}, where the strongest peaks at $v_r<0$ ($v_r=-155$~km\,s$^{-1}$ and -285~km\,s$^{-1}$) align almost exactly with the peaks of the CR and OLR distributions. We also mark a weaker peak ($v_r=-330$~km\,s$^{-1}$) which is close to the higher resonances, including $l/m=3/2$ (purple). This weaker peak may be the same as Chevron 5 found by \citet{belokurov_chevrons}. These strikingly close matches are strong evidence that the chevrons are indeed generated by resonances rather than a merger event.

The retrograde chevrons do not align with the $m=2$ resonances. The most prominent (Chevron 2, peaking at $r\approx13$~kpc) lies between the $l/m=1$ and $l/m=3/2$ resonances. However, we also show the $l/m=5/4$ resonance (red), which is continuous in energy with the prograde 4:1 ultraharmonic resonance. This does match the chevron well, and their peaks are reasonably well aligned in the bottom two panels of Fig.~\ref{fig:vr_slice} ($v_r=-200$~km\,s$^{-1}$). This chevron being due to an $m=4$ resonance may also explain why it is more symmetric in $v_r$ than the prograde chevrons. When viewed in the $(X,Y)$ frame corotating with the bar, an $m=4$ closed periodic orbit displays four-fold rotational symmetry with pericentres at four different values of $\phi$. This means that the orbit can pass close to the Sun on multiple parts of its orbit when either $v_r>0$ or $v_r<0$, unlike the $m=2$ orbits in Fig.~\ref{fig:orbits_vr}. The radial phase space chevrons would therefore not be dominated by stars with $v_r<0$.

A pair of peaks at $v_r=-285$~km\,s$^{-1}$ and $v_r=-310$~km\,s$^{-1}$ also align very closely with the $l/m=7/4$ and $l/m=2$ resonances respectively. However, we note that the orbits of higher resonant orbits have larger apocentres and are therefore more sensitive to the accuracy of the Galactic potential at larger radii. Since we are only considering one potential and pattern speed it is not possible to confidently associate these higher resonances with individual chevrons.

In summary, the most prominent prograde chevrons can be associated with the CR and OLR, while the clearest retrograde chevron matches the $l/m=5/4$ resonance.

\subsection{Asymmetry in radial phase space}

As Figs.~\ref{fig:r_vr_model} and \ref{fig:r_vr_sim} show, asymmetry in $v_r$ in radial phase space is characteristic of resonance-generated chevrons when a volume-limiting selection effect is applied to the observations. We now directly compare the asymmetry of the models and the data. Following a similar method to \citet{donlon2023}, we calculate the normalised antisymmetric component of the $(r,v_r)$ distributions. This is defined as
\begin{equation}\label{eq:r_vr_antisym}
    f_\mathrm{anti}(r,v_r)\equiv\frac{n(r,v_r)-n(r,-v_r)}{n(r,v_r)+n(r,-v_r)},
\end{equation}
where $n(r,v_r)$ is the number of stars in the 2D histogram bin located at $(r,v_r)$. This antisymmetric component is shown for the two models and the data in Fig.~\ref{fig:r_vr_antisym}. Red (blue) indicates an excess (deficit) of stars compared to the opposite value of $v_r$.

For prograde stars the asymmetry previously discussed is seen as red pixels at $v_r<0$ for $r\lesssim13$~kpc. Both models and the data all show this effect, although we note that the corotation resonance (peaking at $r\approx10$~kpc) is considerably fainter than the other resonances in the simulation. The outer chevron in the data (associated with the OLR) cannot be traced at $r\gtrsim13$~kpc, due to low signal-to-noise. However, both the analytic model and the simulation make a prediction at these larger radii. As was seen in Figs.~\ref{fig:r_vr_model} and \ref{fig:r_vr_sim}, the asymmetry changes sign at $r\approx13-16$~kpc, with more stars in the chevrons at $v_r>0$.  This can be understood by looking at the orbits in Fig.~\ref{fig:orbits_vr}. While it is not clear from our sample whether this occurs in the Milky Way, future observations may allow this prediction to be tested.

The isochrone model and simulation both predict similar effects in the retrograde stars, with the asymmetry changing sign at $r\approx13$~kpc. However, the retrograde chevrons in the observed data do not show any significant asymmetry. \citet{donlon2023} explained this as evidence that they result from an ancient merger whose debris has been well phase mixed. As discussed in Section~\ref{section:data_resonances}, the most prominent retrograde chevron aligns reasonably well with the predicted $l/m=5/4$ resonance, whose orbits would be expected to be more symmetric in $v_r$ than the $m=2$ resonances. The retrograde observations can therefore be explained by resonances if the $m=4$ rather than $m=2$ resonances are dominant. We note however that this does not appear to be the case in the simulation.

\subsection{Asymmetry in configuration space}

Fig.~\ref{fig:r_vr_antisym} shows that the asymmetry in $v_r$ changes as a function of Galactic radius $r$. However, it must also change as a function of azimuth $\phi$. In particular, Fig.~\ref{fig:orbits_vr} implies that the asymmetry changes sign if the observer's position is reflected in the major or minor axes of the bar. In Fig.~\ref{fig:mean_vr_position} we show the mean $v_r$ of stars with $|J_\phi|<500$~kpc~km\,s$^{-1}$ in bins of cylindrical radius $R$ (up to $R=18$~kpc) and azimuth $\phi$. As in Fig.~\ref{fig:r_vr_antisym}, the columns show the analytic model, simulation and data, and the rows show prograde and retrograde stars. The bar is marked in grey, and the position of the Sun is marked with the symbol $\odot$.

The panels for the isochrone model are simply a binned version of Fig.~\ref{fig:orbits_vr}, so show the behaviour described above. For both prograde and retrograde stars, the mean $v_r$ is negative near the Sun and changes sign when moving either to larger $R$ or across the major axis of the bar. The same effect is also seen in the simulation. We note however that the mean $v_r$ is not perfectly antisymmetric about the bar's major axis: it is closer to zero on the opposite side from the Sun (i.e. on the leading edge of the bar). This is likely to be because the bar in the simulation is slowing with time, resulting in it lagging behind the trapped orbits in azimuth.

For prograde stars the same effect appears to be visible in the data, especially the flip in asymmetry when crossing the major axis. There is therefore reasonable agreement with the resonance origin scenario. We note that this pattern can also arise if the distances for the farthest stars are overestimated, so this result should be viewed with caution. As discussed in relation to Fig.~\ref{fig:r_vr_data}, it is not clear whether the flip in asymmetry is visible at larger $r$ due to low signal-to-noise. Future distance catalogues may allow a better comparison to be made.

As in the radial phase space in Fig.~\ref{fig:r_vr_antisym}, there is weaker agreement between the data and models for the retrograde stars. While the mean $v_r$ does become positive in the direction of the bar's major axis, this happens nearer the azimuth of the Sun than in the simulation. It is therefore not clear whether the retrograde chevrons are consistent with being bar-generated.

In the data the most prominent chevron is the one we associate with the CR, peaking at $r\approx10$~kpc. We now consider only stars close to this chevron. Similar to the selection of stars in Fig.~\ref{fig:data_resonances}, we choose prograde stars with $|J_\phi|<500$~kpc~km\,s$^{-1}$ and radial actions $J_r$ within 100~kpc~km\,s$^{-1}$ of the corotation resonance (for $\Omegab=35$~km\,s$^{-1}$kpc$^{-1}$). In Fig.~\ref{fig:vr_asymmetry_phi} we show the fraction $f(v_r>0)$ of these stars which have $v_r>0$ in bins of Galactocentric azimuth $\phi$. The top, middle and bottom panels show the isochrone model, simulation and data respectively. The bar's major axis lies at $\phi=\phi_\mathrm{b}$ and the Sun is at $\phi-\phi_\mathrm{b}=-30^\circ$. As a control we also calculate $f(v_r>0)$ for stars in the simulation and data in the complementary cuts (still $|J_\phi|<500$~kpc~km\,s$^{-1}$ but with $|J_r-J_{r,\mathrm{res}}|>100$~kpc~km\,s$^{-1}$).

Both models and the data show the same qualitative trend. As previously shown, less than half the stars near this resonance have $v_r>0$ around the Sun. This reverses on the opposite side of the bar's major axis ($\phi>\phi_\mathrm{b}$) for each model as well as the data. This can be explained by the orientations of the resonant orbits in Fig.~\ref{fig:orbits_vr}. Meanwhile, the simulation and data control samples are more symmetric in $v_r$ at all $\phi$. The innermost chevron in the data is therefore qualitatively consistent with predictions for the corotation resonance as a function of $\phi$.

\section{Conclusions}
\label{section:conclusions}

We present a theory for the origin of the `chevron' overdensities in the radial phase space of low-angular momentum stars in the Milky Way \citep{belokurov_chevrons}. We have proposed \citep{dillamore2023} that the chevrons are populated by stars trapped in the resonances of the Galactic bar. To investigate this hypothesis we have developed an analytic model in the isochrone potential \citep{Henon, ELS}, in which all orbits, integrals of motion and frequencies can be written analytically. Alongside this, we run a test particle simulation of a stellar halo in a realistic Milky Way potential with a slowing rotating bar. We compare the predictions of the model and simulation to observations from \textit{Gaia}, supplemented by distance estimates by \citet{zhang2023}. The principal features of our model and our conclusions are summarised below.

\begin{enumerate}[label=\textbf{(\roman*)}]
    \item The resonant orbits with low angular momentum combine to form a series of nested overdensities in radial phase space $(r,v_r)$. The shapes of these are similar to the chevrons formed by highly radial merger events and to those discovered in the Milky Way by \citet{belokurov_chevrons}.\\[-2mm]

    \item The orientations of the stable $m=2$ resonant orbits with respect to the bar change as a function of angular momentum $J_\phi$. On disc-like orbits at high $J_\phi$, the pericentres of the stable orbits are aligned with the minor axis of the bar. At low angular momentum ($J_\phi\lesssim800$~kpc~km\,s$^{-1}$), the stable orientations rotate by $90^\circ$ and the pericentres become aligned with the bar's major axis.\\[-2mm]
    
    \item The stable resonant orbits at low $J_\phi$ have Galactocentric radial velocity $v_r<0$ when they pass through the Solar neighbourhood. This means that when a sample of stars on highly eccentric orbits is viewed from the Sun, chevrons corresponding to bar resonances are predicted to be considerably more populated at $v_r<0$. This is consistent with what is observed in data from \textit{Gaia} DR3 for prograde stars, but not with chevrons arising from an ancient merger event \citep{donlon2023}.\\[-2mm]

    \item In action space $(J_\phi, J_r)$, the resonances lie along diagonal lines from high $J_r$ at $J_\phi=0$ to $J_r=0$ at high $|J_\phi|$. Our analytic model and simulation predict that as $J_\phi$ is increased from zero, the mean $v_r$ transitions from negative to positive along each of these bands. We find that such features are also visible in the data when the action space is colour-coded by mean $v_r$. At a bar pattern speed of $\Omegab=35$~km\,s$^{-1}$ kpc$^{-1}$ consistent with previous studies \citep[e.g.][]{Sa19,chiba2021_treering, Cl22}, these features align reasonably well with the corotation (CR), outer Lindblad (OLR) and 1:1 resonances. The band we associate with the CR is continuous with the Hercules stream in the disc (at low $J_r$). This band extends as far as $J_\phi=0$ and transitions to a mean $v_r<0$ at $J_\phi\lesssim800$.\\[-2mm]

    \item At $\Omegab=35$~km\,s$^{-1}$kpc$^{-1}$, we select stars from the data close to the predicted resonances. The distributions of these stars in radial phase space align almost exactly with the most prominent observed chevrons. In particular, the stars near CR closely match the innermost prograde chevron \citep[`Chevron 1' in][]{belokurov_chevrons} which peaks at $r\approx10$~kpc. This chevron can therefore be seen as a highly eccentric counterpart of the Hercules stream. The next prograde chevron (Chevron 3) corresponds to the OLR. The clearest retrograde chevron (Chevron 2) does not match any of the predicted $m=2$ resonances. However, it does reasonably match the $l/m=5/4$ resonance. This may explain its lack of asymmetry.\\[-2mm]

    \item We predict that the asymmetry in $v_r$ should change as a function of position in the Galaxy. More resonant stars are expected to have $v_r$ at larger radii ($r\gtrsim13$~kpc) and on the opposite sides of the bar's major or minor axes. For stars near the CR, we find that this flip in asymmetry does appear to occur in the data on the opposite side of the bar's major axis. Future data samples may allow these predictions to be tested more comprehensively.\\[-4mm]
\end{enumerate}

Our results show that the observed radial phase space of the Milky Way's halo closely matches many aspects of bar resonances in the models. Various features of the observed chevrons do not appear to be consistent with formation by an ancient merger event, such as the presence of the chevrons at high metallicity  and the asymmetry in $v_r$ \citep{belokurov_chevrons}. These observations are natural consequences of our dynamical explanation.

The local distribution of stars shows (e.g. Fig.~\ref{fig:E_Lz_mean_vr}) clear evidence for a lack of phase mixing. Asymmetry in $v_r$ can help to identify resonant substructures and to distinguish them from the phase-mixed debris of ancient merger events, which should be more uniformly distributed in radial angle space. This emphasises the importance of looking at substructures in angle space as well as action space.

We have shown that the prograde chevrons match the predicted locations of resonances, and can be directly related to moving groups in the disc widely associated with bar resonances. We conclude that there is a high probability that much of the substructure in $(r,v_r)$ space results from trapping of halo stars by the bar. This means that the chevrons should not be used to infer the Milky Way's assembly history without extreme caution. However, it does offer possible avenues for future work to constrain properties of the Galactic bar, such as pattern speed and scale length.

\section*{Acknowledgements}
We thank the Cambridge Streams group and attendees of the Galactic Bars 2023 conference for helpful comments and suggestions during this study. We are grateful to the anonymous referee whose comments have helped to improve this manuscript. AMD thanks the Science and Technology Facilities Council (STFC) for a PhD studentship. VB and NWE acknowledge support from the Leverhulme Research Project Grant RPG-2021-205: "The Faint Universe Made Visible with Machine Learning".

This work has made use of data from the European Space Agency (ESA) mission
{\it Gaia} (\url{https://www.cosmos.esa.int/gaia}), processed by the {\it Gaia} Data Processing and Analysis Consortium (DPAC,
\url{https://www.cosmos.esa.int/web/gaia/dpac/consortium}). Funding for the DPAC has been provided by national institutions, in particular the institutions participating in the {\it Gaia} Multilateral Agreement.

This research made use of Astropy,\footnote{http://www.astropy.org} a community-developed core Python package for Astronomy \citep{astropy:2013, astropy:2018}. This work was funded by UKRI grant 2604986. For the purpose of open access, the author has applied a Creative Commons Attribution (CC BY) licence to any Author Accepted Manuscript version arising.

\section*{Data Availability}

This study uses publicly available \textit{Gaia} data. Code used to calculate resonant orbits in the isochrone potential can be found at \url{https://github.com/adllmr/isochrone_resonances}.



\bibliographystyle{mnras}
\bibliography{refs} 




\appendix

\section{Orbits in the isochrone potential}\label{section:orbits_appendix}
The angle variables $\theta_i$ in the isochrone potential can be expressed analytically in terms of the positions $(r,\theta,\phi)$. We here present these transformations as given by \citet{binney_tremaine}. They are expressed in terms of the variable $\eta$, defined by
\begin{equation}
    s=2+\frac{c}{b}(1-e\,\mathrm{cos}\,\eta),
\end{equation}
where
\begin{align}
    c&\equiv\frac{GM}{-2H_0}-b,\\
    e^2&\equiv1-\frac{L^2}{GMc}\left(1+\frac{b}{c}\right),\\
    s&\equiv1+\sqrt{1+r^2/b^2}.
\end{align}
The angles are then
\begin{align}
    \theta_r&=\eta-\frac{ec}{c+b}\mathrm{sin}\,\eta,\\
    \theta_\theta&=\psi+\frac{\Omega_\theta}{\Omega_r}\theta_r-\mathrm{tan}^{-1}\left(\sqrt{\frac{1+e}{1-e}}\mathrm{tan}\left(\tfrac{1}{2}\eta\right)\right)\\
    &\qquad-\frac{1}{\sqrt{1+4GMb/L^2}}\mathrm{tan}^{-1}\left(\sqrt{\frac{1+e+2b/c}{1-e+2b/c}}\mathrm{tan}\left(\tfrac{1}{2}\eta\right)\right),\notag\\
    \theta_\phi&=\Omega+\mathrm{sgn}(J_\phi)\theta_\theta.
\end{align}
Here $\Omega$ is not a frequency, but instead is the longitude of the ascending node. This is the value of $\phi$ at which the particle crosses the $\theta=\pi/2$ plane with $\dot{\theta}<0$ (i.e. $\dot{z}>0$). $\psi$ is the angle measured in the plane of the orbit from the ascending node to the current position of the particle. In the case of orbits confined to the $\theta=\pi/2$ plane, we may set $\Omega=0$ and hence $\psi=\mathrm{sign}(J_\phi)\phi$ without loss of generality.

In practice we generate orbits by creating a grid in $\eta$, and calculate $\theta_r$ from this. By cubic interpolation we then find $\eta$ as a function of a uniform grid in $\theta_r$ (i.e. equally spaced points in time along each orbit). All other quantities are then calculated.

\section{Calculation of $G$ in the isochrone potential}\label{section:G_appendix}

The pendulum equation~\ref{eq:pendulum} contains the quantity $G$, defined by equation~\ref{eq:G}. This can be written as \citep{chiba2021}:
\begin{align}
    G&=\frac{\partial}{\partial\Js}(m\Omega_\phi+l\Omega_r)\\
    &=\left(m\frac{\partial}{\partial J_\phi}+l\frac{\partial}{\partial J_r}\right)(m\Omega_\phi+l\Omega_r)\\
    &=m^2\frac{\partial\Omega_\phi}{\partial J_\phi}+ml\left(\frac{\partial\Omega_\phi}{\partial J_r}+\frac{\partial\Omega_r}{\partial J_\phi}\right)+l^2\frac{\partial\Omega_r}{\partial J_r}.
\end{align}
We wish to analytically calculate $G$ in the isochrone potential. As before we consider orbits confined to the $\theta=\pi/2$ plane, so that we may set $L=|J_\phi|$. It is convenient to define the parameter
\begin{equation}
    \alpha\equiv\frac{1}{2}\left(1+\frac{L}{\sqrt{L^2+4GMb}}\right).
\end{equation}
The derivatives of the frequencies (equations~\ref{eq:Omega_r} and \ref{eq:Omega_phi}) with respect to the actions can then be written as
\begin{align}
    \frac{\partial\Omega_r}{\partial J_r}&=-3\Omega_r\left[J_r+\frac{\alpha}{2\alpha-1}L\right]^{-1},\label{eq:dOmegar_dJr}\\
    \frac{\partial\Omega_r}{\partial J_\phi}&=\frac{\partial\Omega_\phi}{\partial J_r}=\alpha\,\mathrm{sgn}(J_\phi)\frac{\partial\Omega_r}{\partial J_r},\\
    \frac{\partial\Omega_\phi}{\partial J_\phi}&=\alpha^2\frac{\partial\Omega_r}{\partial J_r}+\frac{\partial\alpha}{\partial J_\phi}\mathrm{sgn}(J_\phi)\Omega_r\\
    &=\alpha^2\frac{\partial\Omega_r}{\partial J_r}+2\alpha(1-\alpha)(2\alpha-1)\frac{\Omega_r}{L}.
\end{align}
Therefore $G$ can be expressed as
\begin{equation}
    G=\left[l+m\alpha\,\mathrm{sgn}(J_\phi)\right]^2\frac{\partial\Omega_r}{\partial J_r}+2m^2\alpha(1-\alpha)(2\alpha-1)\frac{\Omega_r}{L}.\label{eq:G_analytic}
\end{equation}
We can evaluate this expression as a function of $J_\phi$ along the resonances using equations~\ref{eq:Omega_r}, \ref{eq:Jr_Jphi_res} and \ref{eq:dOmegar_dJr}. An equivalent expression is given in \citet{Ea96}.


\section{Test particle simulation setup}\label{section:sim_appendix}

In our simulations the relative strength of the non-axisymmetric multipole components representing the bar evolves according to
\begin{align}
    \beta(t)&=\begin{cases}
    \frac{3}{16}\xi^5-\frac{5}{8}\xi^3+\frac{15}{16}\xi+\frac{1}{2} & t<t_1 \\\\
    1 & t\geq t_1,
    \end{cases}\\
    \xi&\equiv2\frac{t}{t_1}-1.
\end{align}
This form matches that used by \citet{dehnen2000}. The bar's pattern speed evolves according to
\begin{align}
    \Omega_\mathrm{b}(t)&=\begin{cases}
      \Omegai & t<t_1 \\\\
      \Omegai\left[1+\frac{1}{2}\eta\,\Omegai(t-t_1)^2/(t_2-t_1)\right]^{-1} & t_1\leq t<t_2 \\\\
      \Omega_2 \left[1 + \eta\,\Omega_2(t-t_2)\right]^{-1}& t\geq t_2,
   \end{cases}\\
   \Omega_2&\equiv\Omegai\left[1+\frac{1}{2}\eta\,\Omegai(t_2-t_1)\right]^{-1},
\end{align}
where $\eta\equiv-\dot{\Omega}_\mathrm{b}/\Omegab^2$. We choose $t_1=2$~s~kpc/km~$\approx2$~Gyr and $t_2\approx3$ Gyr. We set the initial pattern speed to $\Omegai=80$~km\,s$^{-1}$kpc$^{-1}$ and the deceleration parameter to $\eta=0.003$.


\bsp	
\label{lastpage}
\end{document}